\documentclass[useAMS, usenatbib]{mn2e}

\usepackage{graphicx}

\newcommand{\vecv}{\mathbf{v}}

\newcommand{\ts}{T_*}
\newcommand{\cs}{c_{s,*}}

\newcommand{\rhos}{\rho_*}
\newcommand{\fE}{f_{\rm E,*}}

\newcommand{\msun}{M_{\odot}}
\newcommand{\lsun}{L_{\odot}}

\newcommand{\fEavg}{\langle f_{\rm E}\rangle}

\newcommand{\ft}{f_{\rm trap}}

\newcommand{\red}[1]{{#1}}

\begin{document}

\title[Simulations of Dusty Winds]{Numerical Simulations of Radiatively-Driven Dusty Winds}

\author[Krumholz \& Thompson]{
Mark R. Krumholz$^1$\thanks{mkrumhol@ucsc.edu} and 
Todd A. Thompson$^2$\thanks{thompson@astronomy.ohio-state.edu}
\\ \\
$^1$Department of Astronomy \& Astrophysics, University of California, Santa 
Cruz, CA 95064 USA\\
$^2$Department of Astronomy and Center for Cosmology \& Astro-Particle Physics, The Ohio State University, Columbus, OH\\43210-1173 USA}

\maketitle

\begin{abstract}
Radiation pressure on dust grains may be an important mechanism in driving winds in a wide variety of astrophysical systems. However, the efficiency of the coupling between the radiation field and the dusty gas is poorly understood in environments characterized by high optical depths like those in ultra-luminous infrared galaxies (ULIRGs) and massive dense star clusters. We present a series of idealized numerical experiments, performed with the radiation-hydrodynamic code \textsc{orion}, in which we study the dynamics of such winds and quantify their properties. We find that, after wind acceleration begins, radiation Rayleigh-Taylor instability forces the gas into a configuration that reduces the rate of momentum transfer from the radiation field to the gas by a factor $\sim 10 - 100$ compared to an estimate based on the optical depth at the base of the atmosphere; instead, the rate of momentum transfer from a driving radiation field of luminosity $L$ to the gas is roughly $L/c$ multiplied by \red{one plus} half the optical depth \red{evaluated using the photospheric temperature}, which is far smaller than the optical depth \red{one would obtain using the interior temperature}. When we apply our results to conditions appropriate to ULIRGs and star clusters, we find that the asymptotic wind momentum flux from such objects should not significantly exceed that carried by the direct radiation field, $L/c$. This result constrains the expected mass loss rates from systems that exceed the Eddington limit to be of order the so-called ``single-scattering" limit, and not significantly higher. We present an approximate fitting formula for the rate of momentum transfer from radiation to dusty gas through which it passes, which is suitable for implementation in sub-grid models of galaxy formation.  Finally, we provide a first map of the column density distribution of gas in a radiatively-driven wind as a function of velocity, and velocity dispersion.
\end{abstract}


\section{Introduction}
\label{sec:intro}

Dusty winds are ubiquitous in astrophysics: they are driven on scales ranging from single stars \citep[e.g.][]{habing96a} to star clusters \citep[e.g.][]{lopez11a} to entire galaxies \citep[e.g.][]{veilleux05a}. The driving mechanisms of these winds are diverse and in some cases uncertain, but one possible mechanism for many of them is the force exerted by radiation interacting with dusty matter. Photons moving through dusty gas can be scattered or absorbed by dust grains, transferring some of their momentum. The grains, in turn, transfer this momentum to the gas either through hydrodynamic drag or via magnetic fields, possibly giving rise to a wind.  Radiation pressure on dust has been suggested as an important feedback mechanism in regulating star formation on the scales of galaxies \citep{scoville03a,thompson05a,andrews11a} and individual massive stars clusters \citep{odell67a,scoville01a,krumholz09d, fall10a, murray10a, krumholz10b}, for driving dusty fountain flows in normal spirals \citep{chiao72a,elmegreen83a,ferrara93a}, and for driving galaxy-wide superwinds \citep{murray05a,murray11a,hopkins12b}.

However, assessing these claims has been difficult due to limited understanding of the radiation-matter interaction that drives the flow. For optically thin flows the problem is relatively simple, since the state of the radiation field is decoupled from the gas. For optically thick media, however, the problem is significantly more difficult, because the gas is capable of reshaping the radiation field. This can lead to a number of complex phenomena, such as photon bubbles \citep{blaes03a} and radiation Rayleigh-Taylor instability \citep{krumholz09c, jacquet11a, jiang13a}. The existence of these behaviors makes it non-trivial to calculate from first principles whether an object with a given set of properties can produce a radiatively-driven dusty wind at all, and, if it does, what properties that wind is likely to possess. This problem has thus far prevented definitive identification of the driving mechanisms for winds observed in a variety of systems \citep[e.g.][]{faucher-giguere12a, newman12b}.

In \citet[hereafter Paper I]{krumholz12c} we addressed the first part of this problem: under what circumstances do we expect an object to launch a radiatively-driven dusty wind? We developed an idealized model system that allowed us to extract the important dimensionless numbers governing wind launching, and we then conducted numerical experiments with the radiation-hydrodynamics code \textsc{orion} to explore the non-linear behavior of the system. The major results of Paper I are that the behavior of gravitationally-confined, dusty columns of matter subjected to radiative fluxes are governed primarily by two characteristic values: the dust optical depth and the Eddington ratio, both computed using the opacities that apply at the surface (i.e.~the photosphere) of the dusty gas. For high optical depths and surface Eddington ratios close to but below unity, which may describe many galaxies and star clusters, we showed that radiation passing through the gas drives statistically steady turbulence with average Eddington ratio of unity, but not a wind.

This result, however, does not answer the question of what happens if a wind \textit{is} launched -- either because the surface Eddington ratio exceeds unity, or because some other mechanism is able to eject matter, by itself or in conjunction with radiation forces. This question is the main focus of our paper. We seek to determine at what rate the matter in a radiatively-driven wind is able to extract momentum from the radiation field, and how this depends on properties such as the strength of the radiative driving and the optical depth of the matter. In addition to illuminating the physics of the winds, we also derive a rough fitting formula that can be used in numerical simulations that do not include radiation-hydrodynamics, and instead treat radiative driving using sub-grid semi-analytic models \citep[e.g.][]{oppenheimer06a, hopkins11a}. \citet{jiang13a} performed preliminary work on this problem in the context of winds where the dominant opacity source is Thompson scattering from free electrons, and concluded that radiation Rayleigh-Taylor instability would limit the wind mass and momentum flux. We seek to investigate whether the same is true for dusty winds, and to extend their results by drawing quantitative rather than qualitative conclusions about how the wind momentum depends on the properties of the system.

The remainder of this paper is as follows. In Section \ref{sec:model} we briefly review the basic equations and model system developed in Paper I, and consider how to extend them to the case of a dusty wind. In Section \ref{sec:sim} we describe our numerical simulations, and in Section \ref{sec:results} we analyze the results they produce. In Section \ref{sec:discussion} we discuss the implications of our results and provide some caveats, and Section \ref{sec:summary} summarizes our conclusions.

\section{Model System}
\label{sec:model}

\subsection{Governing Equations and Model System}

As in Paper I, we treat a section of a galactic disk or a young star cluster as an idealized model system consisting of a slab of gas with total surface density $\Sigma$ filling the domain $z>0$. A vertical radiation flux $F = F_0 \hat{z}$ enters the domain of interest at $z=0$, and there are no radiation sources at $z>0$ other than the thermal emission of the gas. The slab of material is confined by a constant vertical gravitational force per unit mass $-g\hat{z}$; we neglect the self-gravity of the gas.

Since we are interested in cases where the gas layer is optically thick, we describe this system using the two-temperature flux-limited diffusion (2TFLD) approximation, in which we assume that the radiation spectrum is locally a Planck function at every point, but we do not require that the temperature $T_r$ describing this Planck function be identical to the gas temperature $T_g$. \red{We discuss the limitations of the 2TFLD method in Section \ref{sec:fld}.} In this approximation, interaction of radiation and matter is governed by the Planck and Rosseland mean opacities $\kappa_P$ and $\kappa_R$.

The equations governing this system are \citep{krumholz07b}
\begin{eqnarray}
\label{eq:continuity}
\frac{\partial}{\partial t} \rho & = & -\nabla \cdot (\rho \vecv) \\
\label{eq:momentum}
\frac{\partial}{\partial t}(\rho \vecv) & = & -\nabla \cdot (\rho\vecv\vecv) - \nabla P - \lambda \nabla E - \rho g \hat{z} \\
\frac{\partial}{\partial t}(\rho e) & = & -\nabla \cdot [(\rho e+P)\vecv] - \kappa_P \rho (4\pi B - c E) 
\nonumber \\
& & {} + \lambda \left(2\frac{\kappa_P}{\kappa_R} - 1\right) \vecv \cdot \nabla E - \frac{3-R_2}{2} \kappa_P\rho  \frac{v^2}{c} E 
\nonumber \\
& & {} - \rho g v_z
\end{eqnarray}
\begin{eqnarray}
\frac{\partial}{\partial t}E & = & \nabla\cdot \left(\frac{c\lambda}{\kappa_R \rho} \nabla E\right) + \kappa_P \rho (4\pi B - c E)
\nonumber \\
& & {} - \lambda \left(2\frac{\kappa_P}{\kappa_R}-1\right) \vecv\cdot\nabla E
\nonumber \\
& & {} + \frac{3-R_2}{2} \kappa_P \rho \frac{v^2}{c} E - \nabla \cdot \left(\frac{3-R_2}{2}\vecv E\right),
\label{eq:radenergy}
\end{eqnarray}
where $\vecv$ is the gas velocity, $P = \rho k_B T_g/\mu m_{\rm H}$ is the gas pressure, $\mu$ is the mean molecular weight in hydrogen masses, $e = P/[(\gamma-1)\rho] + v^2/2$ is the gas specific energy, $\mu$ is the mean mass per gas particle in units of the hydrogen mass $m_{\rm H}$, $\gamma$ is the gas ratio of specific heats, $B=c a T_g^4/4\pi$ is the frequency-integrated Planck function, $E = a T_r^4$ is the radiation energy density, $\mathbf{F}$ is the radiation flux, $\lambda$ is the flux limiter, and $R_2$ is the Eddington factor. In the 2TFLD approximation, adopting the flux-limiter of \citet{levermore81a} and \citet{levermore84a}, the radiation quantities are related by
\begin{eqnarray}
\label{eq:fld}
\mathbf{F} & = & -\frac{c \lambda}{\kappa_R \rho} \nabla E \\
\lambda & = & \frac{1}{R} \left(\mbox{coth}\, R - \frac{1}{R}\right) \\
R & = & \frac{|\nabla E|}{\kappa_R \rho E} \\
R_2 & = & \lambda + \lambda^2 R.
\end{eqnarray}
In this work we adopt opacities
\begin{equation}
\label{eq:opacity}
(\kappa_R, \kappa_P) = (10^{-3/2}, 10^{-1}) \left(\frac{T_g}{10\mbox{ K}}\right)^2\mbox{ cm}^2\mbox{ g}^{-1},
\end{equation}
an approximation to the behavior of dust opacity at temperatures below $\sim 150$ K \citep{semenov03a}. \red{For simplicity, as in Paper I, we continue to use this powerlaw approximation even at higher temperatures, and note that this will generally lead us to overestimate the opacity and thus the strength of matter-radiation coupling.}

\red{We pause here to note that a few subtle physical assumptions built into the above equations. First, we have not distinguished between the temperatures of dust and gas, and thereby have implicitly assumed that they are the same. Second, the opacity should depend on the radiation spectrum and thus on the effective radiation temperature; in writing down the opacity law, equation (\ref{eq:opacity}), in terms of the gas temperature, we have implicitly assumed that the gas temperature and the radiation temperature are also tightly coupled, at least in the regime where the opacity is large enough to matter. Third, we have also assumed that dust and gas are dynamically coupled. We show in Appendix \ref{sec:assumptions} that these assumptions are well-justified in the regime we consider.
}

In Paper I we show that these equations are fully characterized by four non-dimensional parameters
\begin{eqnarray}
\fE & = & \frac{\kappa_{R,*} F_0}{gc}\\
\beta_s & = & \frac{\cs}{c} = \frac{1}{c} \sqrt{\frac{k_B}{\mu}} \left(\frac{g}{a \kappa_{R,*}} \fE\right)^{1/8} \\
\tau_* & = & \Sigma \kappa_{R,*} \\
k_0 & = & \frac{\kappa_{P,*}}{\kappa_{R,*}}.
\end{eqnarray}
Here $\kappa_{R,*} = \kappa_R(\rhos,\ts)$ and similarly for $\kappa_{P,*}$, and the starred quantities in turn are defined by
\begin{equation}
\label{eq:nondim1}
T_* = \left(\frac{F_0}{ca}\right)^{1/4} \qquad
\rho_* = \frac{\Sigma g \mu m_{\rm H}}{k_B T_*}.
\end{equation}
We may think of $T_*$ as the characteristic temperature at the photosphere of the dusty gas, where the radiation escapes to infinity. The natural units of velocity, length, and time for the problem are
\begin{equation}
c_{s,*} = \sqrt{\frac{k_B T_*}{\mu m_{\rm H}}}
\qquad
h_* = \frac{c_{s,*}^2}{g}
\qquad
t_* = \frac{h_*}{c_{s,*}}.
\end{equation}
In real astrophysical systems, $\beta_s$ is always very small for any non-relativistic flow, and $k_0$ is always of order unity and probably varies little from one galaxy to another. Thus in practice the quantities $\fE$ and $\tau_*$ determine the behavior of the system.

\subsection{Dimensionless Numbers for Winds}

If gas has been launched into a wind, it has obviously overcome its initial gravitational confinement. As discussed in the Introduction, in Paper I we show that this occurs only for $\fE > 1$, i.e.~only if the Eddington ratio at the dust photosphere exceeds unity. In principle one wishes to determine the properties of winds launched at a range of $\fE>1$. However, we focus on the asymptotic limit $\fE \rightarrow\infty$, corresponding to a freely-accelerating wind with negligible gravitational confinement. Our reasons for doing so are threefold. First, this reduces the parameter space we must explore. Second, it is very likely that the case where there is no gravitational confinement will produce the largest possible wind momentum flux, and so it can serve as a useful upper limit. We will see below that even this upper limit is quite restrictive on the possible momentum of the wind. Third, in the case of radiating optically thick disks (e.g.~starburst galaxy disks or the disks around QSOs), $\fE$ rises with height above the disk, so winds at large distances will have larger $\fE$ values \citep{zhang12a}.

For $\fE\rightarrow \infty$, or equivalently $g\rightarrow 0$, the quantities $\rho_*$, $h_*$, and $t_*$ cease to be well-defined. It is therefore helpful to define alternative natural units in the freely-accelerating wind case. The sound speed $c_{s,*}$ remains the natural unit of velocity, and to define a unit of time it is helpful to ask how long it would take the momentum carried by the direct radiation field to accelerate matter from rest to this speed. The momentum flux per unit mass of the injected radiation field is
\begin{equation}
f_{\rm rad,dir} = \frac{F_0}{\Sigma c},
\end{equation}
and so we define the acceleration time as
\begin{equation}
\label{eq:nondim2}
t_a = \frac{c_{s,*}}{f_{\rm rad,dir}} =  \frac{\tau_* c_{s,*}}{\kappa_{R,*} F_0/c} = \frac{\tau_*}{\fE}t_*.
\end{equation}
If radiative trapping is significant, we expect the matter to increase its velocity by $c_{s,*}$ in a time significantly shorter than $t_a$. Finally, we can define characteristic length and density scales from the combination of $t_a$ and $c_{s,*}$. These are
\begin{equation}
h_a = c_{s,*} t_a = \frac{\tau_*}{\fE} h_* \qquad\quad
\rho_a = \frac{\Sigma}{h_a} =  \frac{\fE}{\tau_*} \rho_*.
\label{eq:nondim3}
\end{equation}
We report all results in this paper in units of $\rho_a$, $h_a$, and $t_a$.

\section{Numerical Simulations}
\label{sec:sim}

\begin{table}
\caption{Simulation Physical Paramters}
\label{tab:simphys}
\begin{tabular}{ccccccc}
\hline
Name &
$\tau_*$ &
$\Sigma$ &
$t_*$ &
$t_a$ &
$h_a/10^{-2}$ &
$\rho_a/10^{-16}$  \\
& 
&
(g cm$^{-2}$) &
(kyr) &
(kyr) &
(pc) &
(g cm$^{-3}$)
\\
\hline
T3 & 3 & 1.4 & 1.1 & 6.9 & 0.38 & 1.2  \\
T10 & 10 & 4.6 & 1.1 & 23 & 1.3 & 1.2 \\
T30 & 30 & 14 & 1.1 & 69 & 3.8 & 1.2 \\ \hline
\end{tabular}
\medskip

Note that T10 describes both runs T10LR and T10HR, which have identical physical parameters but different resolutions and box size.
All models have $T_* = 82$ K, $c_{s,*} = 0.54$ km s$^{-1}$.
\end{table}

\begin{table}
\caption{Simulation Numerical Parameters}
\label{tab:simnum}
\begin{tabular}{cccccc}
\hline
Name &
IC &
$N_x \times N_z$ &
$L_x \times L_z $ &
$\Delta x$ &
$t_{\rm run}$ \\
\hline
T3 & T3F0.5 & $1024\times 16384$ & $85.3 \times 1365$ & $0.083$ & 28.0  \\
T10LR & T10F0.5 & $512\times 32768$ & $25.6 \times 407$ & $0.05$ & 21.9 \\
T10HR & T10F0.5 & $1024\times 16384$ & $25.6 \times 407$ & $0.025$ & 10.6 \\
T30 & - & $1024\times 16384$ & $27.3\times 437$ & $0.027$ & 3.6 \\
\hline
\end{tabular}
\medskip

The initial condition (IC) column gives the name of the corresponding run in Paper I used to produce the initial condition. $N_x\times N_z$ is the size of the computational domain in cells, $L_x\times L_z$ is the size of the computational domain, $\Delta x$ is the size of a computational cell, and $t_{\rm run}$ is the duration for which we run the simulation. All quantities are given in units of $h_a$ and $t_a$. Note that all models were run during the phase where gravity was turned on with $\Delta x/h_* = 0.5$. Finally, for a description of the initial conditions for run T30, see the Appendix.
\end{table}

We solve Equations (\ref{eq:continuity}) -- (\ref{eq:radenergy}) with $g=0$ using the radiation-hydrodynamics code \textsc{orion}. Our simulations are two-dimensional, and take place in the $(x,z)$ plane; a flux $F$ of radiation is injected at the bottom of the computational box, $z=0$. The boundary conditions are periodic in the $x$ direction, impermeable at the lower $z$ boundary and open at the upper $z$ boundary. More details on the boundary conditions are given in Paper I. All other parameters of the simulations are also the same as in Paper I, except that we set the external gravitational field $g=0$, so that $\fE \rightarrow \infty$.

To study the behavior of winds in the limit $\fE\rightarrow \infty$ as a function of $\tau_*$, we run four simulations, which we denote T3, T10LR, T10HR, and T30; the first of these have $\tau_* = 3$, the second two have $\tau_* = 10$, and the third has $\tau_* = 30$. The two $\tau_* = 10$ runs are identical except in their resolutions and the sizes of the computational domains. We perform both runs as a check on the resolution-dependence of our results. We summarize the physical and numerical parameters of the simulations in Tables \ref{tab:simphys} and \ref{tab:simnum}.

In Paper I, we show that initially laminar, gravitationally-confined gas slabs subjected to radiation fluxes such that $\fE$ is above a certain critical value, but less than unity, will develop radiation-driven turbulence. This occurs in a time $\la 100t*$. (Note that values of $t_*$ for our runs are given in Table \ref{tab:simphys}, and are the same for every run; $100t_* \approx 100$ kyr.) Since this is short compared to most astrophysical time scales of relevance, it is reasonable to assume that gas being launched in a wind will be in a fully turbulent state. We therefore do not start our simulations with laminar gas layers. Instead, we use as initial conditions the end states of the simulations from Paper I, as summarized in Table \ref{tab:simnum}. We modify these conditions only in that we place the gas in a computational box that is larger in the vertical direction, in order to accommodate vertical expansion of the gas layer once it is no longer gravitationally-confined. We initialize computational cells that are outside the computational domain of the simulations of Paper I by giving them densities equal to the background density and temperature from Paper I, $10^{-10}\rho_*$ and $T_*$, respectively. In run T10LR, we also down-sample the resolution by a factor of 2. The exception to the above statements is run T30, for which we do not have a corresponding run from Paper I. We describe how we generate its initial conditions in the Appendix.

As the simulations proceed, when necessary we shift all velocities in the computational domain by a constant offset in order to bring the center of mass velocity of the gas back to zero. Our method is simple: we have added an option to the \textsc{orion} code that, upon restart from a checkpoint, calculates the center of mass velocity of the computational domain in the $z$ direction, then subtracts the corresponding velocity from all computational cells, altering the momenta and total energies appropriately. The calculation then restarts from the modified data. We apply this option whenever a visual inspection of the data indicates that the bulk of the mass is well away from the bottom boundary of the computational domain. This enables us to continue the simulations longer without the gas reaching the top of the computational box. We shift the velocities in this manner only when the vast majority of the gas is well away from the bottom of the computational box, so that there are no significant artificial forces exerted by the bottom of the computational box. In the analysis below, we remove these offsets and present the results as if the entire simulation had simply taken place in a larger box. In principle we could shift the positions as well, but this is less convenient computationally, since it would require translating values from one cell to another, and filling in values of density, momentum, and energy in the new cells added to the computational domain by any shifts.

\section{Results}
\label{sec:results}

\subsection{Qualitative Behavior}

\begin{figure}
\begin{center}
\includegraphics[height=8in]{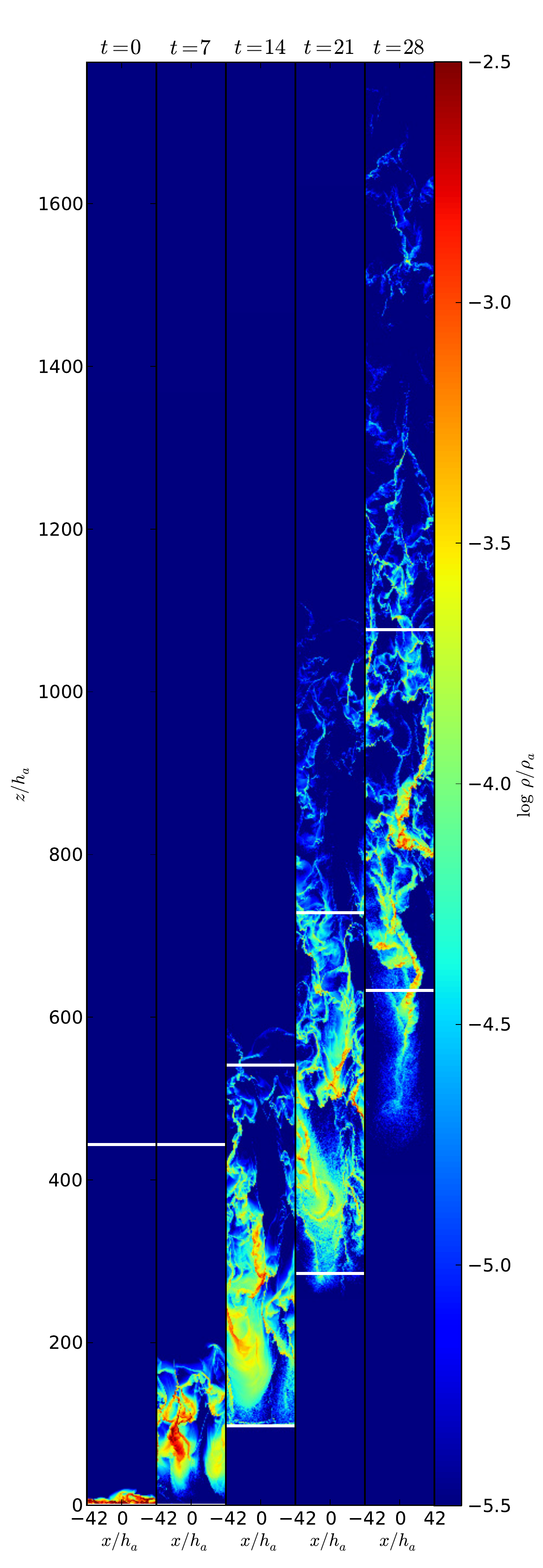}
\end{center}
\caption{
\label{fig:rhoplot_t3_all}
Density distribution as a function of time in run T3. Snapshots are shown at intervals of $5t_a$, starting from $t=0$, as indicated at the top of each panel. White bars indicate a region around the vertical center of mass; we show a zoom-in of this region in Figure \ref{fig:rhoplot_t3_zoom}. Note that the vertical extent shown does not necessarily match the size of the computational box given in Table \ref{tab:simnum}, because we have compensated for the effects of our periodic shifts the center of mass velocity of the entire computational domain in some runs -- see Section \ref{sec:sim} for details.
}
\end{figure}

\begin{figure}
\begin{center}
\includegraphics[height=8in]{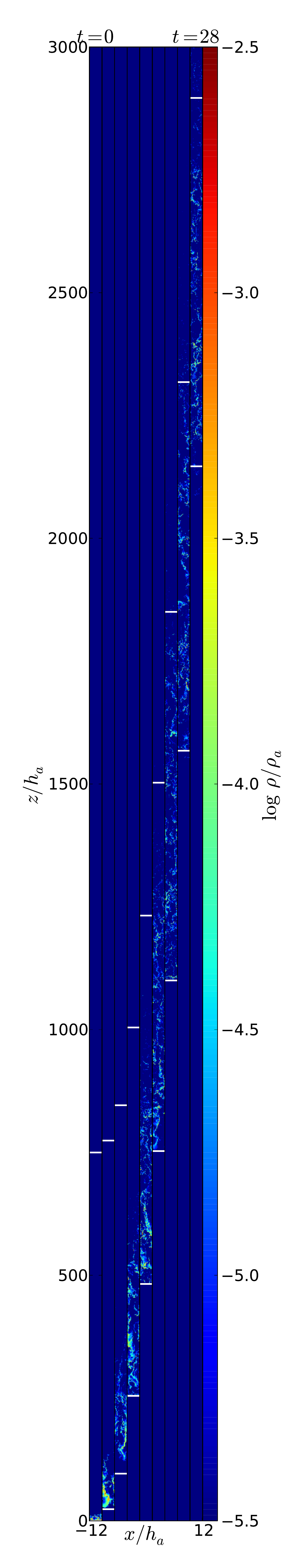}
\end{center}
\caption{
\label{fig:rhoplot_t10lr_all}
Same as Figure \ref{fig:rhoplot_t3_all}, but for run T10LR. The zoomed-in region is shown in Figure \ref{fig:rhoplot_t10lr_zoom}.
}
\end{figure}

\begin{figure}
\begin{center}
\includegraphics[height=8in]{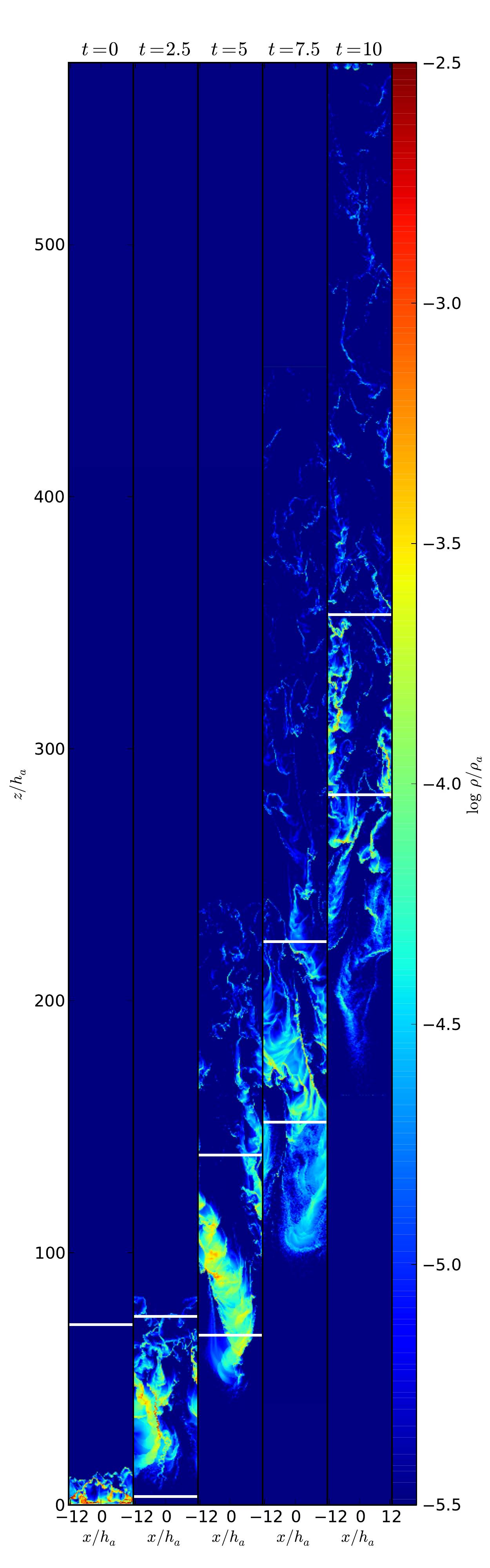}
\end{center}
\caption{
\label{fig:rhoplot_t10hr_all}
Same as Figure \ref{fig:rhoplot_t3_all}, but for run T10HR. The zoomed-in region is shown in Figure \ref{fig:rhoplot_t10hr_zoom}.
}
\end{figure}

\begin{figure}
\begin{center}
\includegraphics[height=8in]{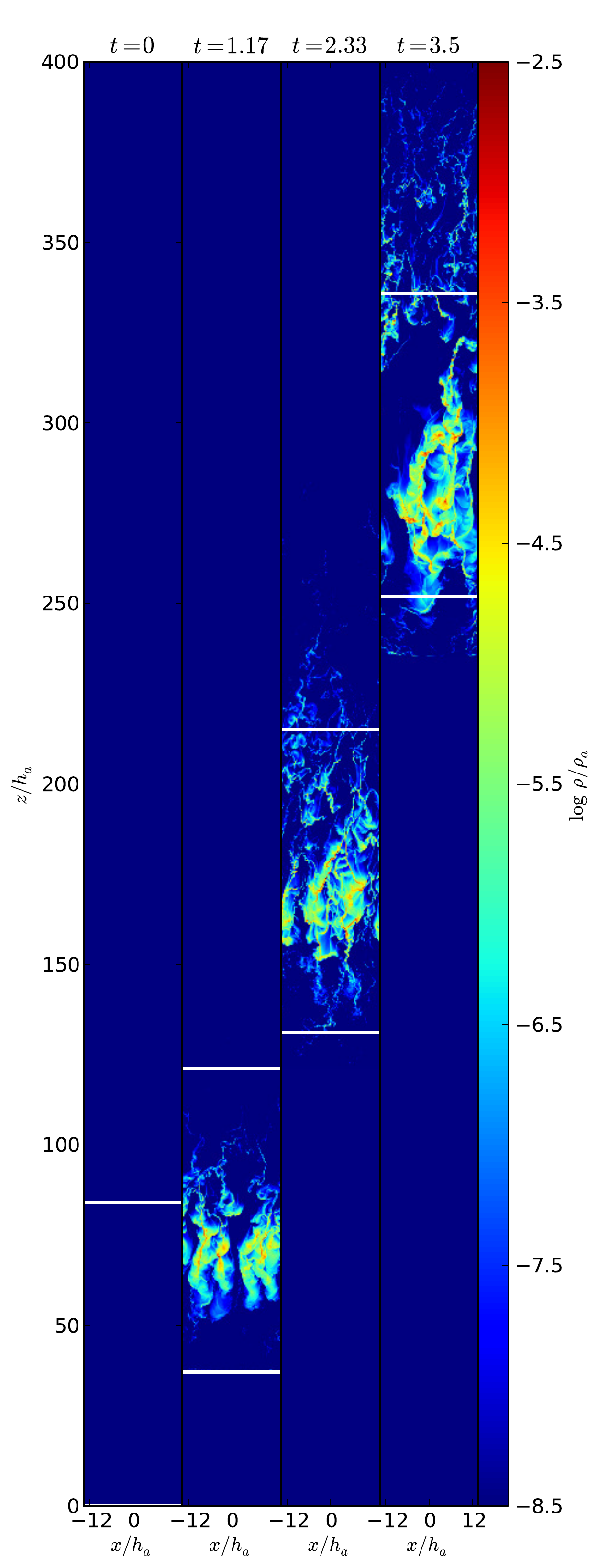}
\end{center}
\caption{
\label{fig:rhoplot_t30_all}
Same as Figure \ref{fig:rhoplot_t3_all}, but for run T30. The zoomed-in region is shown in Figure \ref{fig:rhoplot_t30_zoom}. Note that the first panel is not in fact empty -- the gas at time 0 is simply compressed into an extremely thin layer whose width, on the scale plotted, is less than a single pixel.
}
\end{figure}

\begin{figure}
\includegraphics[width=84mm]{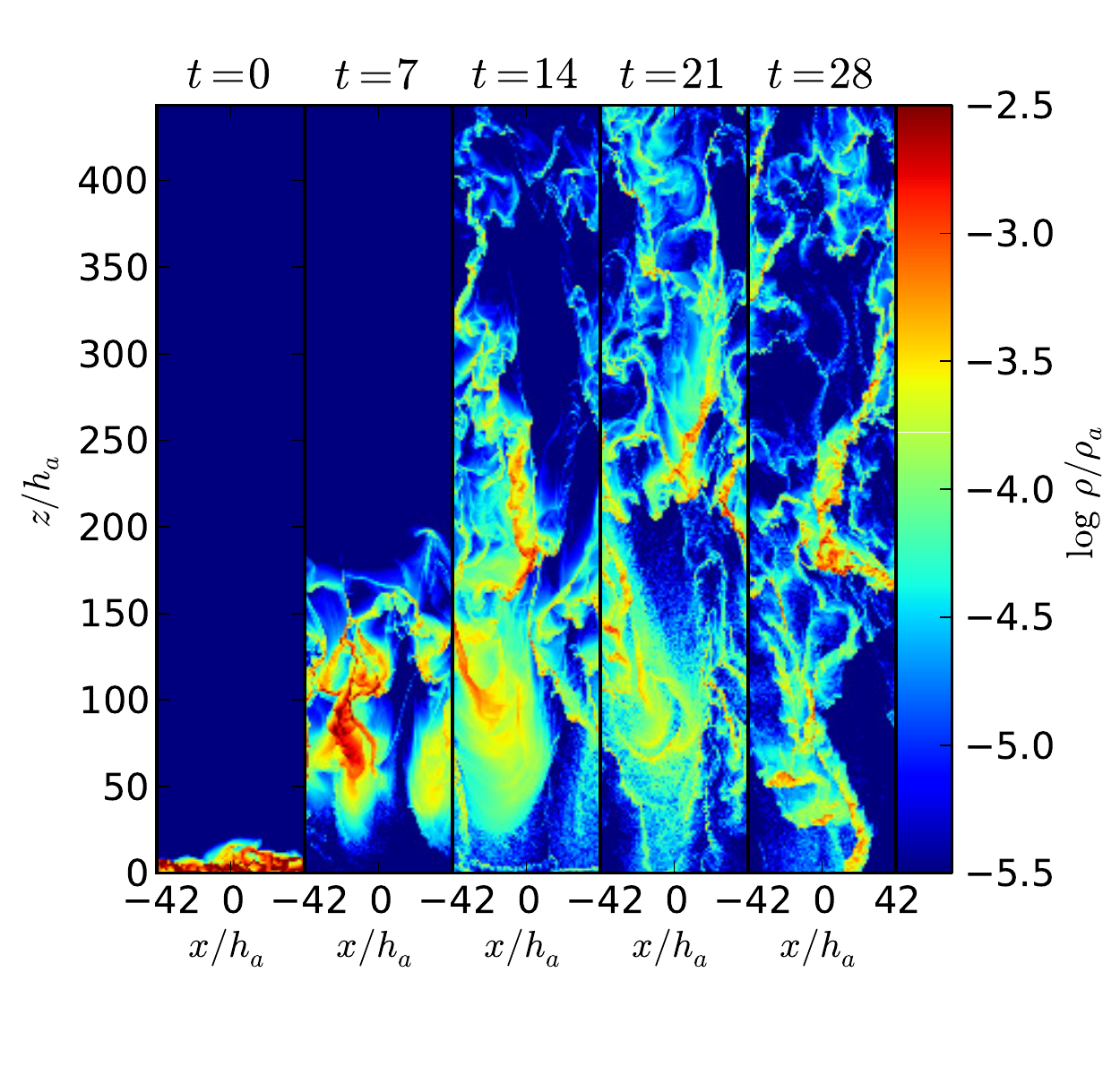}
\caption{
\label{fig:rhoplot_t3_zoom}
Same as Figure \ref{fig:rhoplot_t3_all}, but the panels show a zoom-in on a small region around the vertical center of mass. The zoomed region is indicated by the white bars in Figure \ref{fig:rhoplot_t3_all}.
}
\end{figure}

\begin{figure}
\includegraphics[width=84mm]{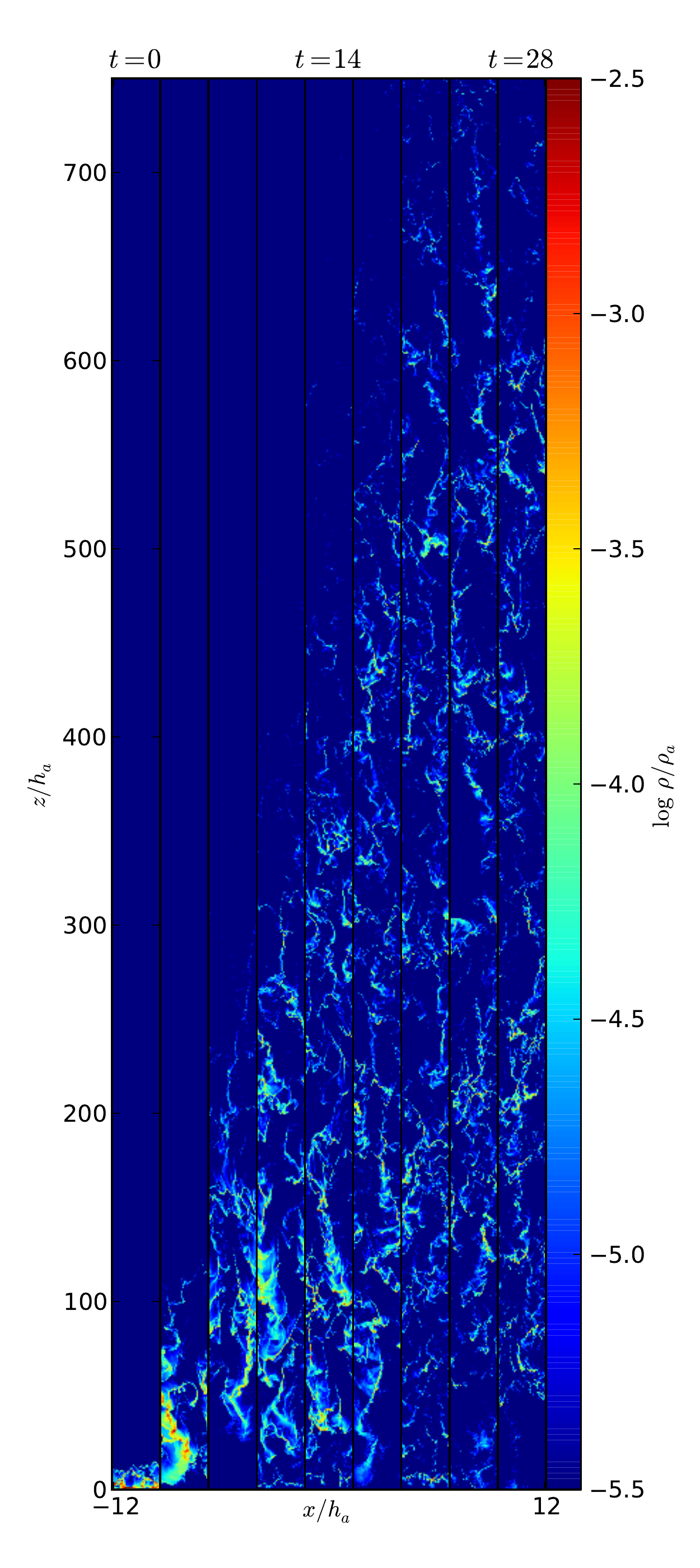}
\caption{
\label{fig:rhoplot_t10lr_zoom}
Same as Figure \ref{fig:rhoplot_t3_zoom}, but for run T10LR.
}
\end{figure}

\begin{figure}
\includegraphics[width=84mm]{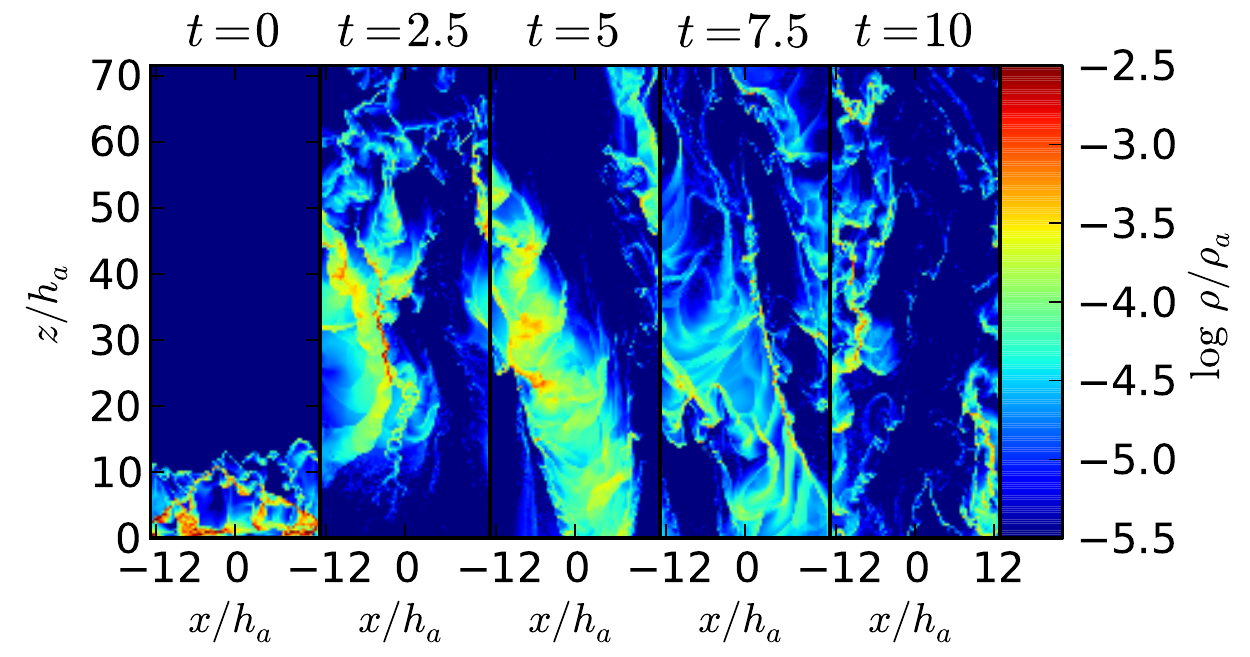}
\caption{
\label{fig:rhoplot_t10hr_zoom}
Same as Figure \ref{fig:rhoplot_t3_zoom}, but for run T10HR.
}
\end{figure}

\begin{figure}
\includegraphics[width=84mm]{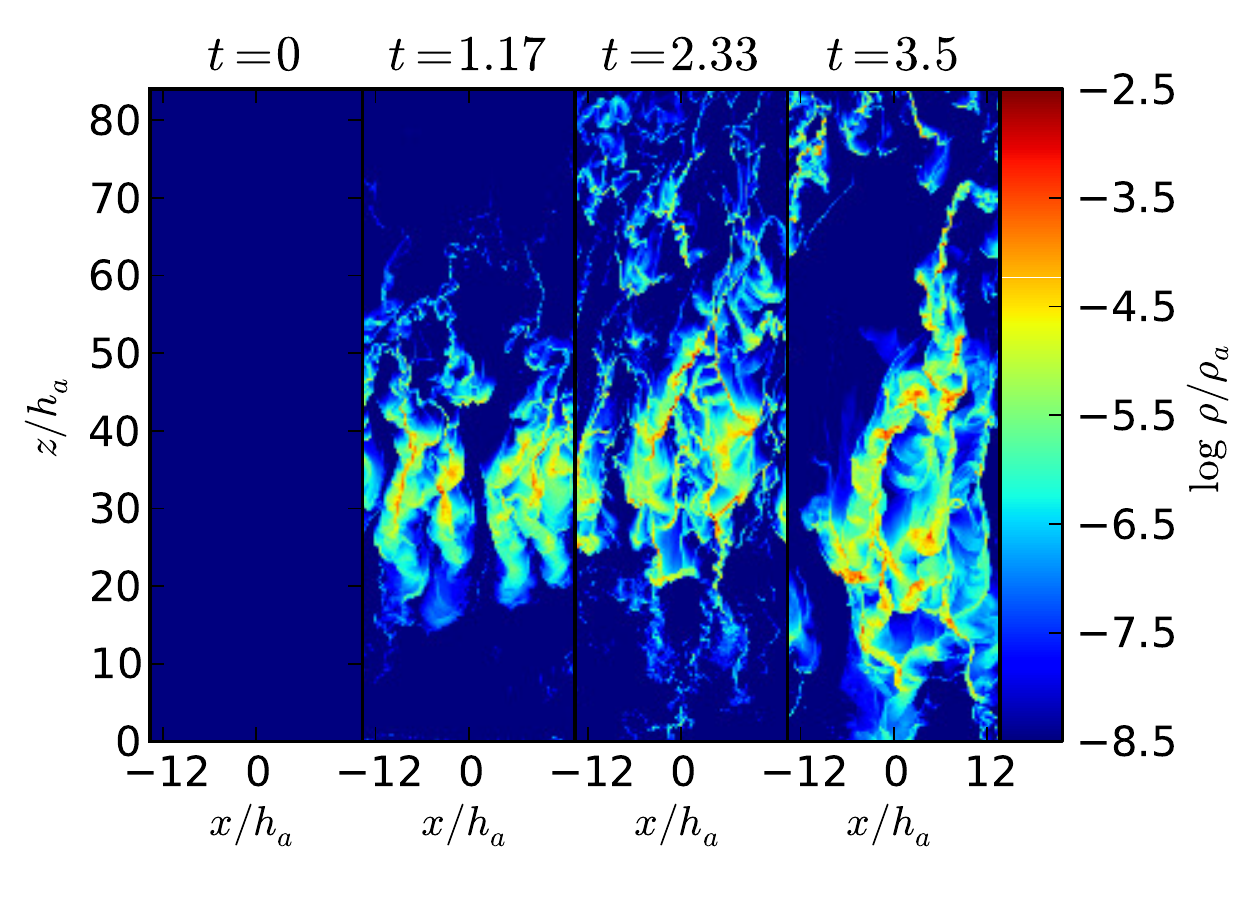}
\caption{
\label{fig:rhoplot_t30_zoom}
Same as Figure \ref{fig:rhoplot_t3_zoom}, but for run T30. As with Figure \ref{fig:rhoplot_t30_all}, note that the first panel is not empty, but it appears so because the gas at time 0 is simply compressed into an extremely thin layer whose width, on the scale plotted, is less than a single pixel.
}
\end{figure}

Figures \ref{fig:rhoplot_t3_all} -- \ref{fig:rhoplot_t30_zoom} shows a series of snapshots of the simulation density fields. As the plots show, the initial state that results from the radiation Rayleigh-Taylor (RRT) instability \citep{jacquet11a} acting on a gas confined by gravity, consists of a relatively horizontal, turbulent layer. In the absence of gravitational confinement, the radiation force rapidly drives the gas into a predominantly vertical, filamentary structure. In between the filaments of dense gas there are low-density channels. As the material is accelerated upward by the radiation field, the gas becomes more elongated and spread over a progressively larger vertical extent. We are eventually forced to halt our simulations primarily because the vertical extent of the gas becomes comparable to the vertical size of our computational domain.

\begin{figure}
\includegraphics[width=84mm]{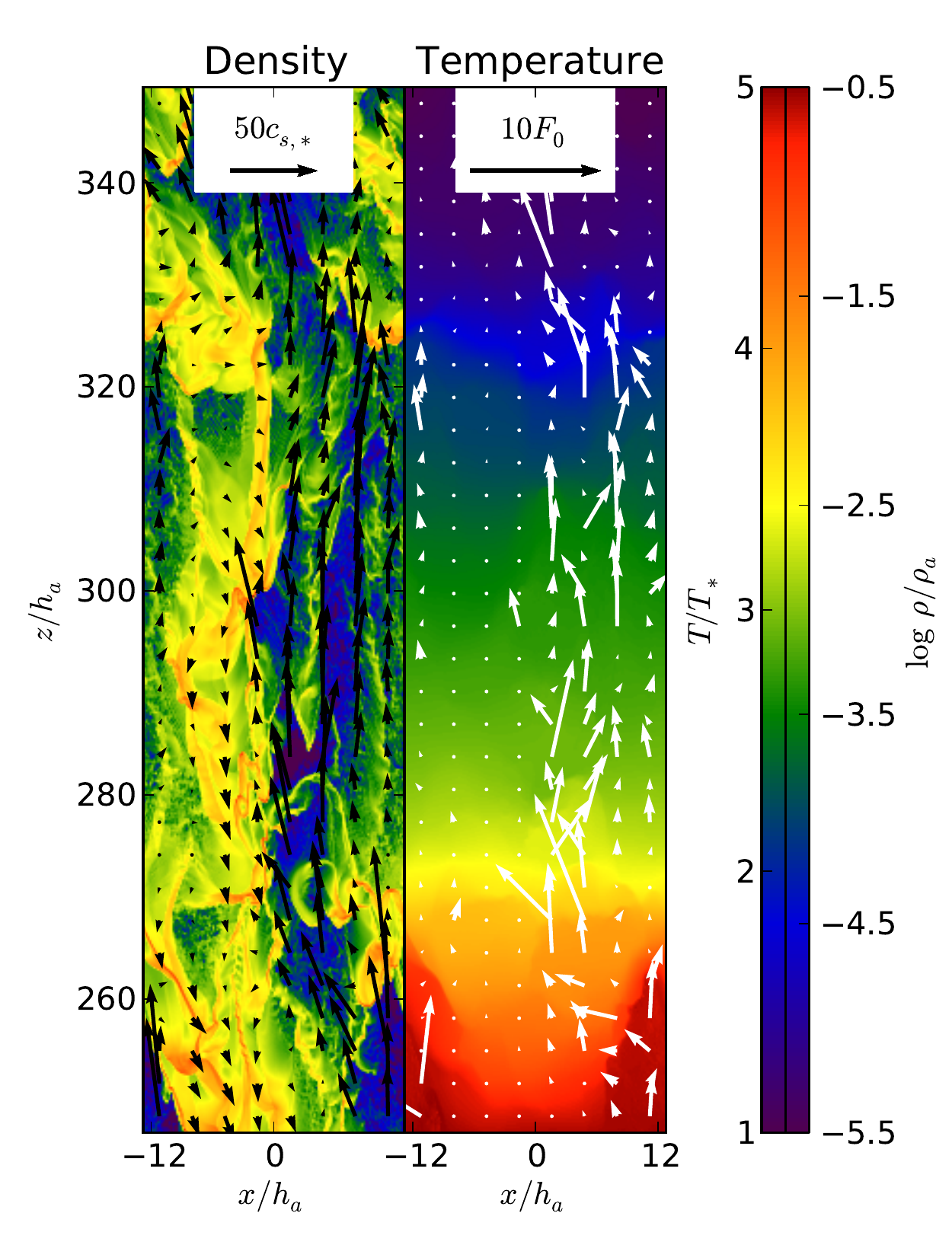}
\caption{
\label{fig:rhotvec}
A section from run T10LR at $t=12.4 t_a$ showing the density and velocity distribution (colors and vectors in the left panel), and the gas temperature and radiative flux distribution (colors and vectors in the right panel). The region shown is centered in the vertical center of mass of the gas at this time, $z_{\rm cm} = 298h_a$, and the velocities shown are relative to the vertical center of mass velocity of the section shown, $v_{z,\rm cm}=77.0 c_{s,*}$.
}
\end{figure}

Figure \ref{fig:rhotvec} shows an example of the distribution of density, temperature, velocity, and radiation flux in one of the runs once the channel structure has developed.  The channels are characterized by several features. First, within them the gas is traveling at extremely large velocities relative the dense gas in the filaments. At the snapshot shown, the velocity difference approaches many tens of $c_{s,*}$. As a result of this velocity difference, the edges of the channels appear to be scalloped by Kelvin-Helmholtz instabilities. Second, because of their lower optical depths, the channels carry the great majority of the radiative flux. The flux within the channels approaches $10 F_0$, while inside the filaments the flux is $\ll F_0$. Thus the matter effectively collimates the radiation field, inducing a strong anti-correlation between density and radiative flux. This anti-correlation is the main signature of the RRT instability.

\subsection{Radiative Trapping}

\begin{figure}
\includegraphics[width=86mm]{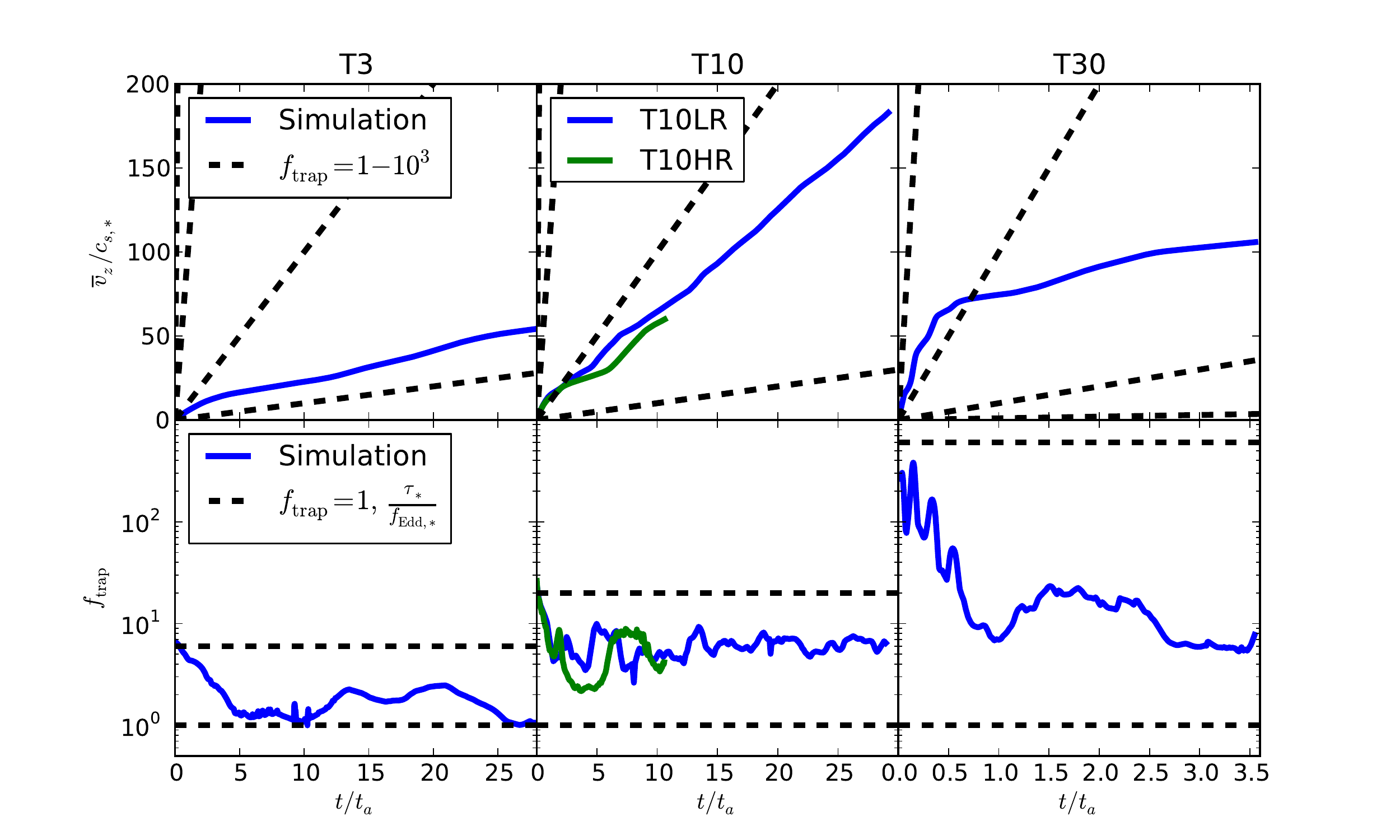}
\caption{
\label{fig:ftrap}
Mean gas velocity versus time (top row) and trapping factor versus time (bottom row) for runs T3 (first column), T10 (second column), and T30 (third column). In all panels solid lines show the results of the simulations; for T10, blue shows T10LR, and green T10HR. In the upper panels black dashed lines show the change in velocity versus time that would be expected for trapping factors $\ft = 1, 10, 100$, and $1000$, from shallowest line to steepest. In the bottom panel, dashed black lines show $\ft =1$ and $\ft = \tau_*/\fE$, with the value of $\fE$ computed \textit{before} gravity is turned off -- i.e.~the value of $\fE$ that was used in the simulation from Paper I from which we take our initial conditions. In the absence of gravity, as is the case for the simulations shown here, $\fE = \infty$.
}
\end{figure}

The development of vertical filamentary structure and the resulting collimation of the radiation field has profound effects on its ability to trap the radiation field and extract momentum from it. To quantify the rate at which the gas takes up momentum from the radiation field, it is helpful to examine the $z$ component of the momentum equation including radiation and gravitational forces; this is
\begin{equation}
\frac{d}{dt}(\rho v_z) = -\nabla\cdot(\rho \vecv v_z) - \frac{dP}{dz} + \kappa_R \rho \frac{F_z}{c} - \rho g,
\end{equation}
where $F_z$ is the $z$-component of the radiation flux and the use of $\kappa_R$ in the equation implicitly equates the flux-mean and Rosseland-mean opacities, as is appropriate in the optically thick regime. In the flux-limited diffusion approximation this equation is equivalent to Equation (\ref{eq:momentum}), as shown by \citet{krumholz07b}, but the analysis is more transparent when the equation is written in the form above. If we integrate this equation over the entire computational domain, and ignore the small terms that arise from forces and fluxes across the top and bottom boundaries of the computational domain, the first two terms on the right-hand side vanish and we are left with
\begin{equation}
\frac{d}{dt} \left\langle \rho v_z\right\rangle = \left\langle \kappa_R \rho \frac{F_z}{c}\right\rangle - \langle\rho\rangle g,
\end{equation}
where for any quantity $q$ we defined the volume average by
\begin{equation}
\langle q \rangle = \frac{1}{L_x L_z} \int_{-L_x/2}^{L_x/2} \int_0^{L_z} q \, dz \, dx.
\end{equation}
Dividing both sides by $\langle\rho\rangle$ and noting that the mass in the computational domain is very close to constant (since we are careful to ensure there is no significant mass loss from the top of the computational box) gives
\begin{equation}
\label{eq:zmom}
\frac{d\overline{v}_z }{dt} = \frac{1}{c} \frac{\langle \kappa_R \rho F_z\rangle}{\langle\rho\rangle} - g,
\end{equation}
where $\overline{v}_z = \langle\rho v_z\rangle/\langle\rho\rangle$ is the mass-weighted mean $z$ velocity of the gas. We use the first term on the right-hand side to define the mean radiation force per unit mass,
\begin{equation}
f_{\rm rad} =  \frac{1}{c} \frac{\langle \kappa_R \rho F_z\rangle}{\langle\rho\rangle}.
\end{equation}
Equivalently, we may think of this term as describing the mass-weighted mean radiation force. Based on our observation that density and flux are strongly anti-correlated, we expect that to be much less than the volume-weighted mean radiation force $\langle \kappa_R F_0\rangle/c$ would be. The second term on the right-hand side is simply the gravitational force per unit mass. 

At this point it is useful to rewrite the equation by multiplying through by a factor of $t_a/c_{s,*}$ to non-dimensionalize. Doing so gives
\begin{equation}
\label{eq:dvdt}
\frac{t_a}{c_{s,*}}\frac{d\overline{v}_z }{dt} = \frac{f_{\rm rad}}{f_{\rm rad,dir}} - \frac{\tau_*}{\fE}.
\end{equation}
Following \citet{krumholz09d} and \citet{krumholz12c}, we define the trapping factor by
\begin{equation}
1 + f_{\rm trap} = \frac{f_{\rm rad}}{f_{\rm rad,dir}}.
\end{equation}
Physically, the trapping factor is simply the factor by which the radiation force is amplified by trapping of the radiation field by the gas. The quantity $1+\ft$ is equivalent to the amplification factor $\tau_{\rm IR}$ defined by \citet{thompson05a}, although we refer to it as $\ft$ here because, as we will see, its relationship to optical depth is not trivial. Using equation (\ref{eq:dvdt}) to rewrite the equation above, we obtain
\begin{equation}
\label{eq:ftrap}
f_{\rm trap} = \frac{t_a}{c_{s,*}}\frac{d\overline{v}_z}{dt}  +  \frac{\tau_*}{\fE} - 1.
\end{equation}
The quantity on the right-hand side is directly measurable from our simulations (and $\fE = \infty$ in the absence of gravity, so the term $\tau_*/\fE = 0$), so our simulations provide us with a direct measurement of $f_{\rm trap}$ as a function of time. However, we must make one important modification to equation (\ref{eq:ftrap}), which comes from a limitation of our numerical method. Because we are using flux-limited diffusion, we do not properly capture the interaction of the gas with the direct, beamed radiation field produced by stars. Instead, we are treating the radiation field only after this first absorption. Since the final $-1$ represents the contribution from this direct radiation field, we do not subtract it off when computing $f_{\rm trap}$ from the simulations. This is likely conservative, since our method does capture some of the effects of the first absorption, in which case the results we obtain should be upper limits on $f_{\rm trap}$. However, we cannot completely rule out the possibility that inclusion of the direct radiation force would somehow change the structure of the gas and indirectly increase the trapping of the reradiated field.

Figure \ref{fig:ftrap} shows the gas mean velocity and trapping factor as a function of time in each of our simulations. For constant $\ft$, the gas velocity should increase linearly with time. Instead, we see that the velocity increase is steep at first and then becomes much shallower, and this is reflected in the plots of $\ft$, which are large at first and then decline over a few $t_a$. The initially high values are easy to understand given our starting conditions. When the gas is confined by gravity and there is no wind, both the time-averaged value of $\overline{v}_z$ and its rate of change must be zero. Consulting equation (\ref{eq:ftrap}), this requires that $\ft = \tau_*/\fE$, again omitting the $-1$ because our simulation does not properly model the direct radiation force. The initial value of $\ft$ we measure is indeed close to this, though the match is not exact because $d\overline{v}_z/dt$ is not precisely zero at all times in the gravity-confined state; instead, it oscillates about zero.

\begin{table*}
\begin{minipage}{126mm}
\caption{Simulation Results}
\label{tab:simresult}
\begin{tabular}{cccccccc}
\hline
Name &
$t_{\rm min}$ &
$\min(\ft)$ &
$\langle\ft\rangle$ &
$\max(\ft)$ &
$\tau_*/\fE$ &
$\kappa(T_{\rm mp})\Sigma$ &
\red{$\langle\tau\rangle$} \\
&
&
&
&
&
$\approx f_{\rm trap,grav}$ &
$\approx \tau_{\rm IR}$ &
\\
\hline
T3 & 5 & 1.0 & $1.6\pm 0.4$ & 2.5 & 6 & 15 & \red{$6.2\pm 0.6$} \\
T10LR & 3 & 2.6 & $6.0\pm 1.3$ & 9.9 & 20 & 120 & \red{$32\pm 3$} \\
T10HR & 3 & 2.2 & $4.9\pm 2.4$ & 8.9 & 20 & 120 & \red{$30\pm 5$} \\
T30 & 1 & 5.3 & $12.3\pm 5.7$ & 23.3 & 600 & 2000 & \red{$132\pm 11$} \\ \hline
\end{tabular}
\medskip

For each run, $\min(\ft)$, $\langle \ft\rangle$, and $\max(\ft)$ give the minimum, time-averaged, and maximum values of $\ft$ that occur in the simulation after time $t_{\rm min}$. The error bars given on $\langle\ft\rangle$ represent the $1\sigma$ range measured from the simulations at times $>t_{\rm min}$. For comparison, $\tau_*/\fE\approx f_{\rm trap,grav}$ gives the time-averaged value of $\ft$ in the steady-state gravitationally-confined configuration from which we start, while $\kappa(T_{\rm mp}) \Sigma \approx \tau_{\rm IR}$ is the average optical depth at the start of the calculation, computed using the mass-weighted mean midplane temperature as in Paper I. \red{The quantity $\langle\tau\rangle$ is the mass-weighted mean optical depth of the computational domain computed using the temperatures from the simulations; see equation (\ref{eq:taudef}). As with $\langle\ft\rangle$, the value shown is the time average at times $>t_{\rm min}$, and the error bars give the $1\sigma$ range over this time.}
\end{minipage}
\end{table*}

Once the gravitational confinement is removed, however, the gas morphology changes from predominantly horizontal to predominantly vertical, and $\ft$ drops. At late times $\ft$ oscillates up and down about a value well below the initial one. Upward and downward oscillations of $\ft$ correspond to variations in the gas morphology. At times, for example at times $t/t_a = 5$ and $10$ in run T10HR (see Figure \ref{fig:rhoplot_t10hr_zoom}), the filaments formed by the radiation are fairly coherent leave fairly large vertical channels unobstructed, and at these times $\ft$ is low. At other times, such as $t/t_a = 7.5$ in run T10HR, the filaments are more fragmented and cover more or the domain horizontally, giving rise to larger values of $\ft$.

Finally, comparing runs T10LR and T10HR suggests that are results are relatively well-converged. Since the gas is turbulent for RRT instability in its fully developed state, the flow is chaotic and we do not expect either morphologies or exact values of $\ft$ as a function of time to be resolution-independent. However, examining the results in Figure \ref{fig:ftrap}, we see that values of $\ft$ versus time produced in the two runs are qualitatively similar, and that quantitatively their means are well within the level of variance in $\ft$ we measure in each run as the flow varies chaotically. This suggests that our values of $\ft$ are converged.

We summarize our results for $\ft$ in Table \ref{tab:simresult}, where we report the minimum, mean, and maximum values of $\ft$ we measure in each of our simulations once the initial transient phase ends. For comparison, we also report $\tau_*/\fE\approx f_{\rm trap,grav}$, the mean value of $\ft$ in the initial, turbulent, gravity-confined state, and $\kappa(T_{\rm mp}) \Sigma \approx \tau_{\rm IR}$, the optical depth computed by multiplying the column density by the opacity evaluated using the midplane temperature $T_{\rm mp}$. The latter has been used as an approximate value for $\ft$ by a number of authors, as we discuss in more detail below. Clearly none of these values are equal; instead $\ft \ll f_{\rm trap,grav} \ll \tau_{\rm IR}$.

\subsection{Wind Velocity Distribution}

\begin{figure}
\includegraphics[width=84mm]{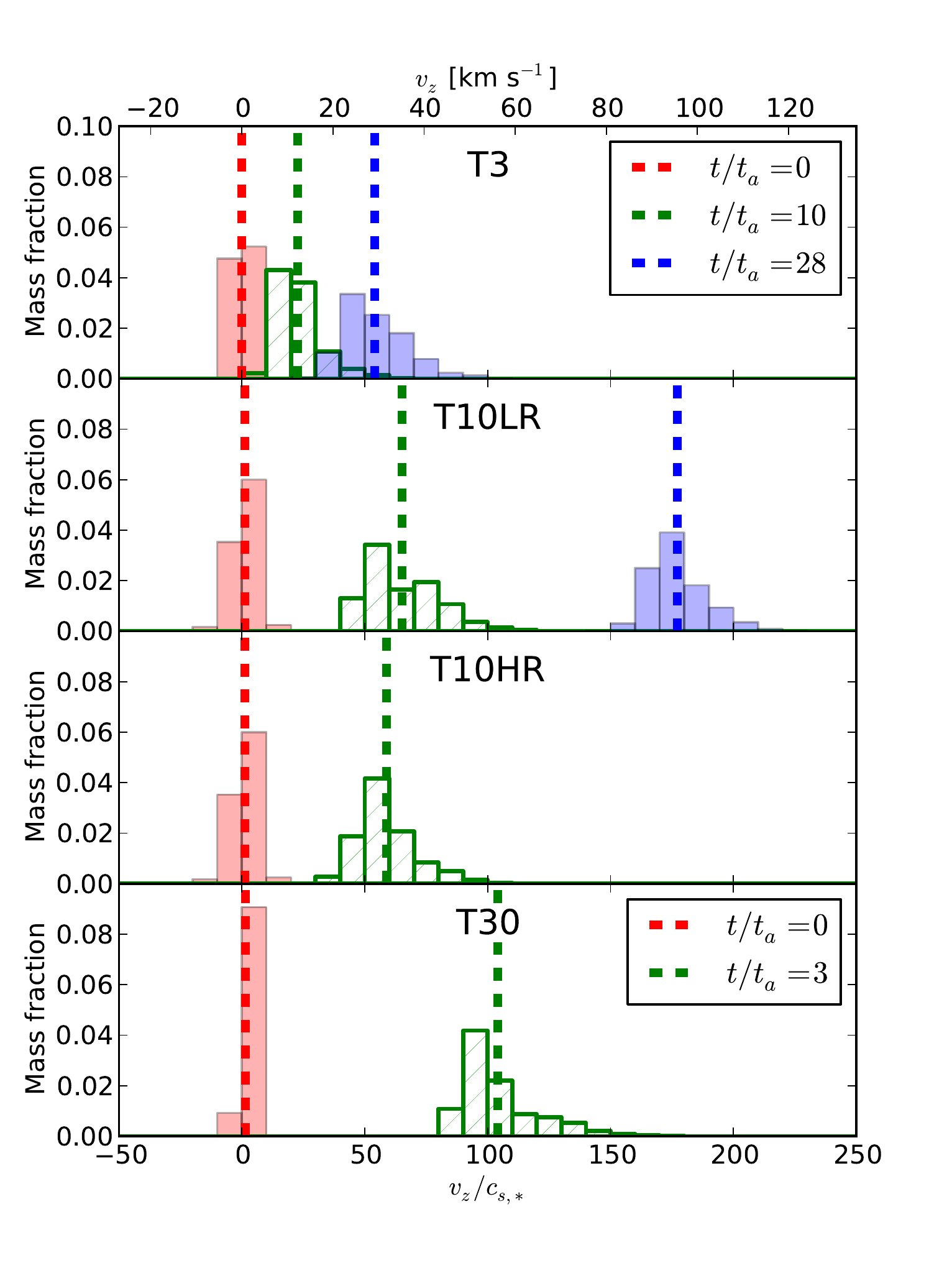}
\caption{
\label{fig:vdist}
Velocity distribution functions for each of our simulations at times $t/t_a = 0$, $10$, and $28$ (for runs T3, T10LR, and T10HR) and $t/t_a = 0$ and $3$ (for run T30), as indicated in the legend. In each panel the histogram shows the fraction of the mass in the simulation that falls into a given bin of $z$ velocity at the indicated time. Vertical dashed lines indicate the mass-weighted mean velocity $\overline{v}_z$ at that time.
}
\end{figure}

Figure \ref{fig:ftrap} and Table \ref{tab:simresult} describe the mean velocity and mass-averaged momentum transfer from radiation to gas. However, it is also interesting to look at the distribution of matter velocities. In Figure \ref{fig:vdist} we show mass-weighted velocity probability distribution functions (PDFs) for each of our runs at several times. In the initial condition, the distribution of $z$ velocities is, as one might expect for a roughly hydrostatic atmosphere, symmetric about $v_z = 0$. The width is fairly small, reflecting the relatively low Mach numbers we obtained for turbulent atmospheres in Paper I. At times $\gg t_a$, not only is the mean velocity $\overline{v}_z$ larger, the spread of velocities is larger as well. At late times the PDF is all the runs is slightly asymmetric, with the majority of the mass residing at velocities slightly below the mean, and a tail extending well above the mean.

\begin{figure}
\includegraphics[width=84mm]{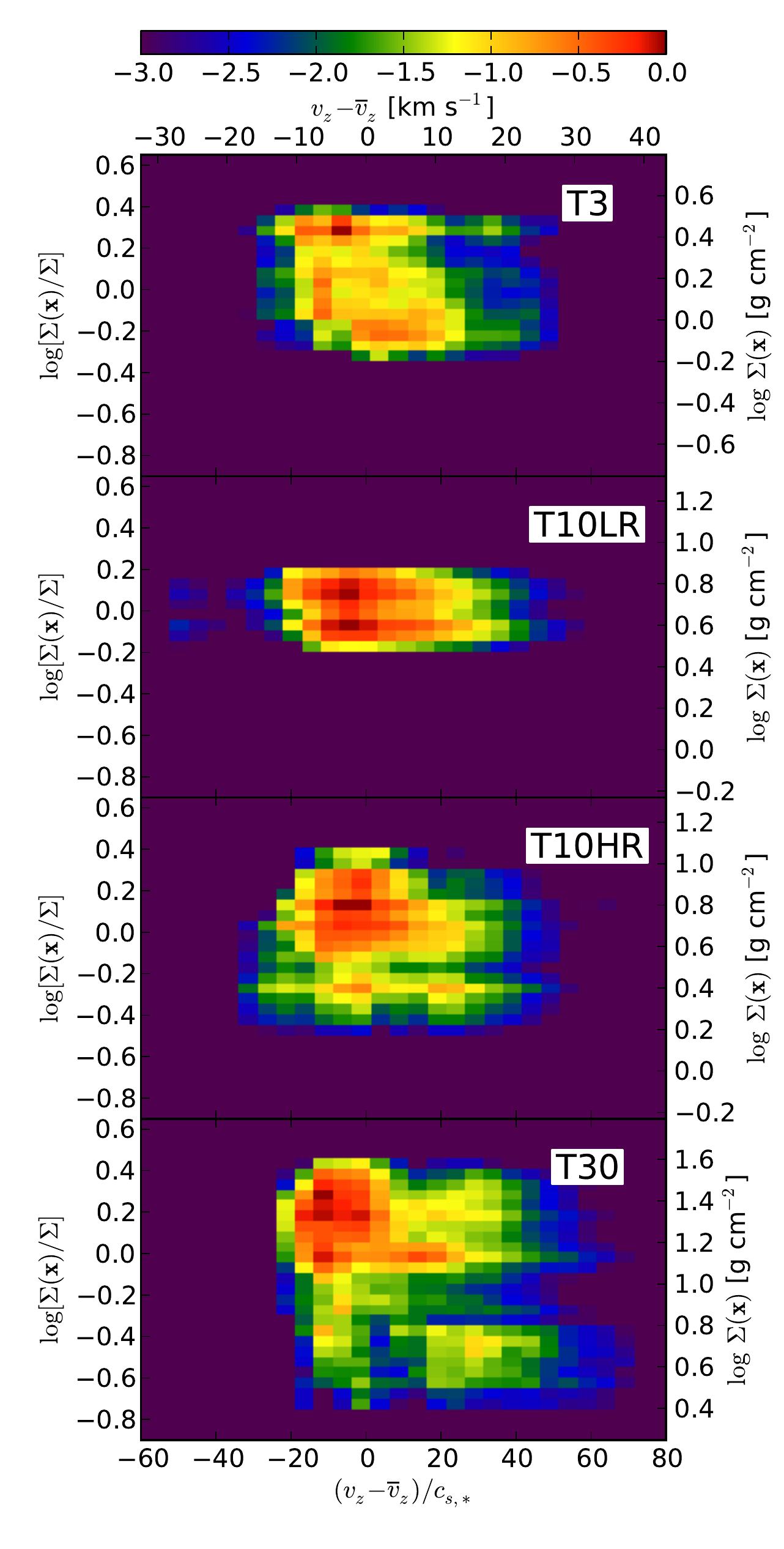}
\caption{
\label{fig:sigmavdist}
Two dimensional velocity-column density distributions in the simulations at the latest time slices shown in Figure \ref{fig:vdist} ($t/t_a = 28$ for runs T3 and T10LR, $t/t_a = 10$ for run T10HR, and $t/t_a = 3$ for run T30). Each pixel shows the logarithm of the fraction of the simulation mass in the indicated bin of $v_z$ and $\Sigma(x)$, normalized so that the most massive bin has a value of unity. Note that both the $x$ and $y$ axes are offset such that material at the mass-weighted mean column density and velocity would appear at the position $(0,0)$.
}
\end{figure}

The division between high and low velocity material corresponds to the division between material in the opaque filaments and material in or at the edges of the radiation-dominated channels, as illustrated in Figure \ref{fig:sigmavdist}. To construct this figure, along every vertical line of sight we measure the column density
\begin{equation}
\Sigma(x) = \int_{0}^{L_z} \rho(x,z) \, dz.
\end{equation}
We then assign every cell a column density $\Sigma(x)$ corresponding to the value at its $x$ position, and construct the two-dimensional PDF of $\Sigma(x)$ and $v_z$. From the 2D PDF, we see the same asymmetry as in Figure \ref{fig:vdist}, where the velocity distribution extends further from the mean in the positive direction than the negative direction. In the 2D PDF, it is clear that the high velocity material consists preferentially of gas with low $\Sigma(x)$. The correlation is relatively weak, and the overall range in $\Sigma(x)$ is relatively small, because the filaments are not perfectly vertical. Thus, most of the time a given vertical line of sight will intersect both dense filaments and low-density channels, rather than looking straight down the barrel of a channel. Nonetheless, this column density-velocity anti-correlation represents a possible observable signature of radiation pressure-driven dusty winds.

The overall width of the velocity distribution, including both low and high speed components, is of order $\sim 20c_{s,*}$ in all the runs; the dispersion of horizontal velocities is substantially smaller. The dispersion does not appear to increase substantially over the time interval shown, and thus $\sim 20c_{s,*}$ is likely the steady-state value, at least over the range of $\tau_*$ values that we have explored. This corresponds to a one-dimensional Mach number in the vertical direction of order 20 -- not exactly 20, since much of the gas is somewhat warmer than $T_*$ and thus has a sound speed greater than $c_{s,*}$. For our fiducial choice of dimensional scaling ($c_{s,*}=0.54$ km s$^{-1}$), this give a physical velocity dispersion of roughly 10 km s$^{-1}$ in the wind, compared to bulk velocities of $\sim 100$ km s$^{-1}$ at the same time. While this Mach number and velocity dispersion are larger than we found in Paper I for RRT-unstable atmospheres that do not drive a wind, they are still close to an order of magnitude smaller than the values observed in the most vigorously star-forming ultraluminous infrared galaxies \citep[e.g.][]{downes98a}. Finally, we note that the results for runs T10LR and T10HR are qualitatively similar at equal times, suggesting at least rough convergence.

\section{Discussion}
\label{sec:discussion}

\subsection{Fitting Formulae for Radiation Trapping}

By combining the results of this paper with those of Paper I, we are now in a position to provide a fitting formula for the value of $\ft$ in optically-thick radiation pressure-driven atmospheres and winds. Such a formula is useful in simulations or analytic calculations that seek to include radiation pressure effects, but that do not properly capture the radiation-hydrodynamic behavior associated with the RRT instability. Examples include one-dimensional models \citep[e.g.][]{krumholz09d, murray10a}, models that include radiation pressure only via a subgrid prescription rather than by solving the equation of radiative transfer \citep[e.g.][]{hopkins11a, agertz12a}, models that solve the transfer equation only for the direct and not the dust-reprocessed radiation field (and therefore implicitly set $f_{\rm trap} = 0$; e.g.~\citealt{peters10b, wise12b, kim13a, kim13b}), and models that solve the transfer equation in one dimension under an assumption of spherical symmetry and therefore miss RRT effects \citep[e.g.][]{novak12a}.

The value of $\ft$ for a radiation pressure-dominated wind or atmosphere is a function of the two main dimensionless parameters for the problem, $\fE$ and $\tau_*$. We have sampled this parameter space quite coarsely, but we can nonetheless provide a rough fit that captures the results of our simulations, and which is an improvement over simple prescriptions. In Paper I we explored the regime $\fE<1$, and found that for a given $\tau_*$ there exists a critical $\fE$ above which instability sets in. At values of $\fE$ below this value the atmosphere is supported predominantly by gas pressure over most of its height, and radiation pressure is dynamically unimportant. Above the critical $\fE$ values, we found that RRT instability causes the value of $\ft$ to self-adjust so that the radiation force exactly balances gravity without producing a wind. This is $\ft\approx \tau_*/\fE - 1$. To extend this to lower $\tau_*$ than we have sampled, we simply impose the requirement that $\ft$ cannot be less than 0. Thus our approximation for $\fE<1$ is
\begin{equation}
f_{\rm trap,lo} \approx \max\left(\frac{\tau_*}{\fE} - 1, 0\right).
\label{eq:ftlo}
\end{equation}

\begin{figure}
\includegraphics[width=84mm]{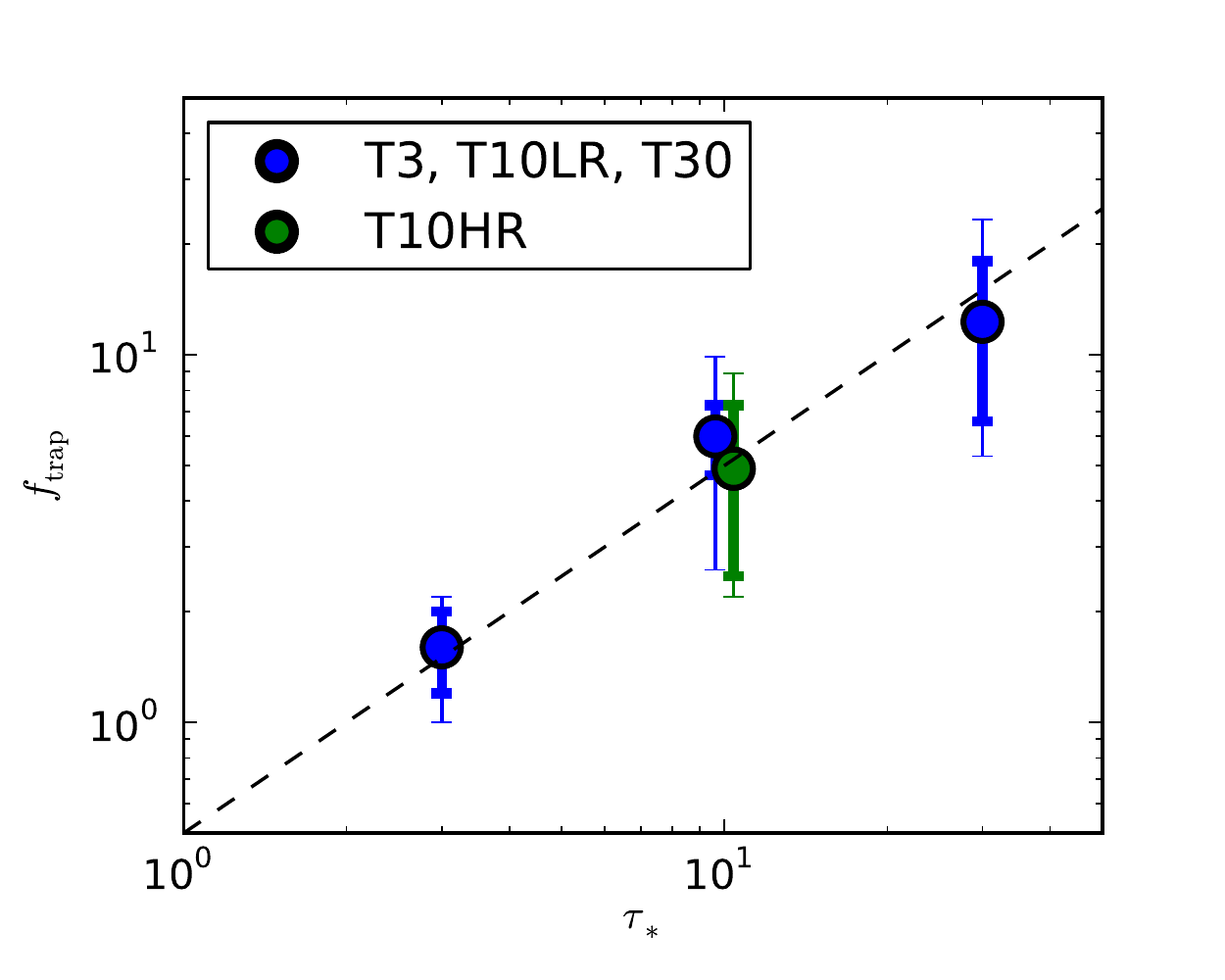}
\caption{
\label{fig:ftraptau}
Values of $\ft$ as a function of $\tau_*$ measured in the simulations presented in this paper. The points represent the time-averaged value, thick error bars show the standard deviation, and thin error bars show the range from minimum to maximum; all values are as given in Table \ref{tab:simresult}. For runs T10LR and T10HR, the points are offset slightly from $\tau_* = 10$ for clarity. The dashed black line is $\ft = 0.5\tau_*$.
}
\end{figure}

In the regime $\fE \rightarrow \infty$ that we explore in this paper there is a wind, but the rate at which it takes up momentum from the radiation field is limited. Figure \ref{fig:ftraptau} shows our estimated values of $\ft$ as a function of $\tau_*$, together with a crude linear fit that is consistent with the simulations:
\begin{equation}
f_{\rm trap,hi} \approx 0.5\tau_*
\label{eq:fthi}
\end{equation}
This is only a ``by-eye" fit, but it describes the data very well, and given the small number of simulations and the error bars on each one, a more sophisticated fitting procedure does not seem justified.

\begin{figure}
\includegraphics[width=84mm]{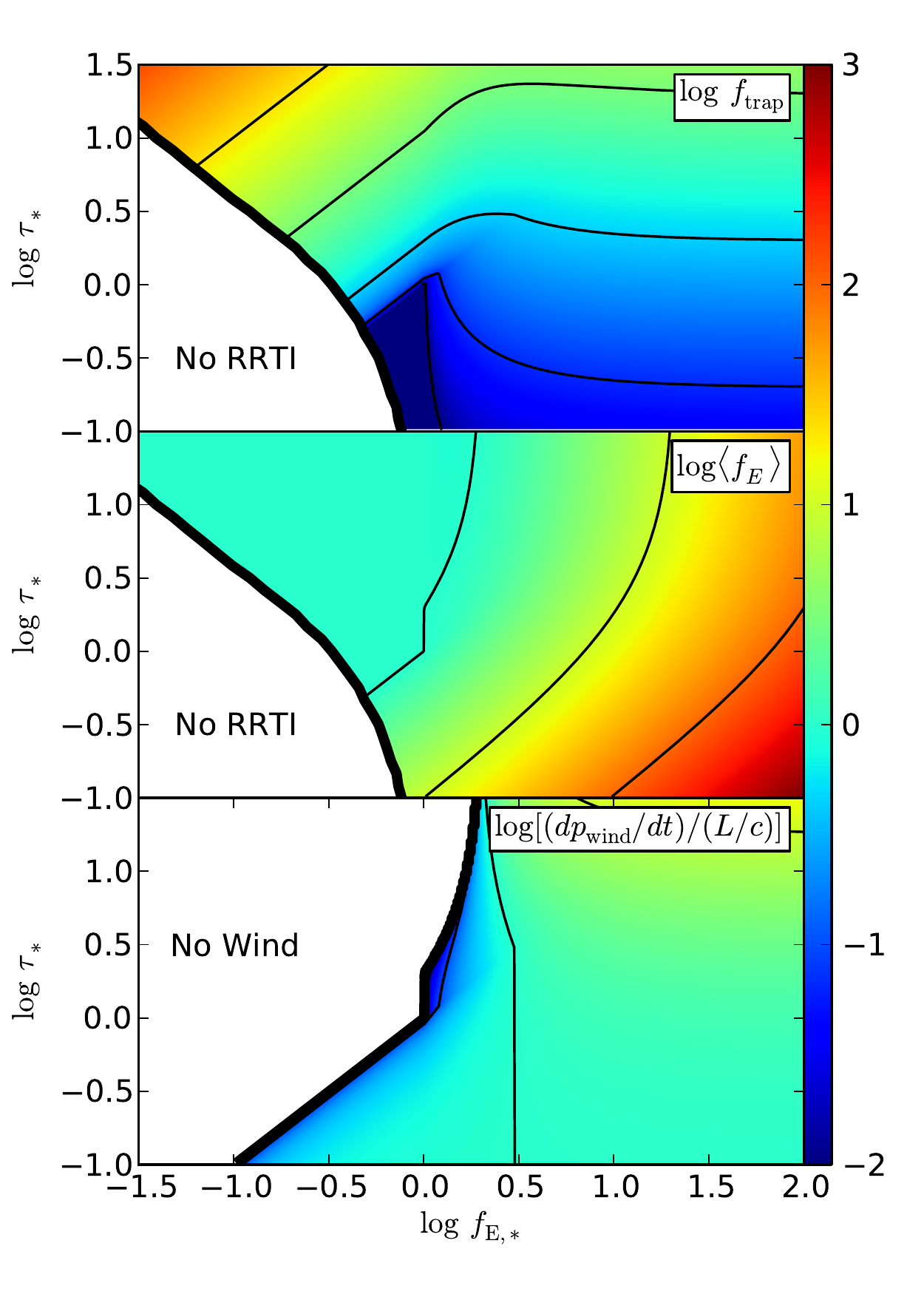}
\caption{
\label{fig:plotfit}
Values of $\log \ft$ (top), $\log \langle f_{\rm E} \rangle$ (middle), and $(dp_{\rm wind}/dt)/(L/c)$ (bottom) as a function of $\fE$ and $\tau_*$, computed using the fitting formula given by Equation (\ref{eq:ftfit}). In each panel contours lines appear at values of $-2,-1,0,1$, and 2. The thick black line in the top two panels shows the critical curve below which RRT instability shuts off (white region). The thick black line in the bottom panel shows the critical value below which no wind is launched.
}
\end{figure}

To combine the two cases $\fE<1$ and $\fE\rightarrow \infty$, we hypothesize that $\ft$ will obey equation (\ref{eq:ftlo}) up to the point where $\fE = 1$. At this point we will have $\ft \approx \tau_* - 1$ for $\tau_* \gg 1$. Beyond this point, as $\fE$ increases $\ft$ will smoothly decrease onto the fit given by equation (\ref{eq:fthi}) in the limit $\fE \rightarrow \infty$. Since we have not mapped out the intermediate $\fE$ regime, obviously the functional form of this decrease is not well-constrained, and we cannot rule out the possibility that $\ft$ behaves non-monotonically over this range, for example developing a peak at a special value of $\fE$. However, there is no good reason to believe that such a phenomenon should occur, and in its absence the
functional form we adopt to interpolate between the behavior at $\fE<1$ and $\fE\rightarrow \infty$ matters little, since $\ft$ only changes by a factor of two between those cases. We therefore adopt a very simple interpolation between the two cases,
\begin{equation}
\ft \approx f_{\rm trap,hi} + \frac{f_{\rm trap,lo} - f_{\rm trap,hi}}{\max(\fE,1)}
\label{eq:ftfit}
\end{equation}
in the regime where RRT instability occurs. This fitting formula has the correct asymptotic behavior in the limits $\tau_*\rightarrow 0$, $\fE < 1$, and $\fE \rightarrow \infty$, and is consistent with the simulations over the range of $\fE$ and $\tau_*$ we have explored. We can also use this formula to determine the ratio of mass-averaged radiation force to gravitational force. In Paper I we showed that this is 
\begin{equation}
\langle f_{\rm E}\rangle = (1 + \ft)\frac{\fE}{\tau_*}. 
\end{equation}
If a wind is launched, from equation (\ref{eq:dvdt}), we can see that the net rate at which it gains momentum including both radiative acceleration and gravitational deceleration is
\begin{equation}
\frac{dp_{\rm wind}}{dt} = \left(1 + \ft\right)\left(\frac{\fEavg-1}{\fEavg}\right) \frac{L}{c}.
\end{equation}

Figure \ref{fig:plotfit} shows the results of our fit for $\ft$, $\langle f_{\rm E}\rangle$, and $(dp_{\rm wind}/dt) / (L/c)$ as a function of $\tau_*$ and $\fE$. In the plot we can see the three dominant regimes identified in our simulations. Below the critical line there is no RRT intstability, and radiation is dynamically subdominant. For $\fE$ large enough to turn on RRTI but still below about unity, $\langle f_{\rm E} \rangle$ is fixed to unity, and $\ft$ self-adjusts to compensate, decreasing as $\fE$ increases at fixed $\tau_*$. Finally, at $\fE$ greater than about unity, a wind appears. In this regime $\ft$ is a function primarily of $\tau_*$, and is quite insensitive to $\fE$. On the other hand $\langle f_{\rm E} \rangle$ increases with $\fE$, indicating that gravity is becoming progressively weaker relative to radiation. As a result, the wind strength is monotonically increasing with $\fE$, but only slowly, since gravity is relatively unimportant once $\fE$ is even a slightly above unity. We stress that the exact location of the wind-launching line should not be taken too seriously, particularly at $\tau_* < 1$, given the sparsity with which we have sampled the parameter space. Nonetheless, the qualitative result that for $\tau_* > 1$ a wind appears only for $\fE \ga 1$ should be robust.

Readers may note that, for $\tau_* \ll 1$, it is possible for there to be a wind even when $\fE < 1$ and RRTI does not occur. Physically, this corresponds to a medium that is optically thin to dust-reprocessed radiation, but is still absorbs the direct radiation field. In this case, the radiation is absorbed once, is reemitted, and then immediately escapes, so the RRTI that we see in our simulations does not occur. However, if the direct radiation field carries enough momentum, this single absorption may still be sufficient to overcome gravity and launch a wind. Simulations by \citet{kuiper12a} suggest that RRTI does not occur in this case, which is not surprising, since RRTI relies upon the ability of the gas to shape the radiation field. That cannot happen if the radiation is only absorbed once.

\subsection{Implications for Star-Forming Systems}

Given our results for $\ft$ and $\fEavg$, it is interesting to ask what our models predict for star-forming systems, which have been posited to be regulated by radiation pressure. We consider two types of objects: proto-star clusters, which we approximate as spherical, and galactic disks, which we approximate as planar. For a spherical object of total (gas plus stellar) mass $M$, gas mass fraction $f_g$, stellar mass fraction $f_*=1-f_g$,  and radius $R$, within which the stars have a light to mass ratio $\Psi$, we have a central luminosity $L=\Psi f_* M$, gas surface density $\Sigma = (1-f_*) M/4 \pi R^2$, surface gravitational acceleration $g = G M/R^2$, surface flux $F = L/4\pi R^2$, and surface temperature $T_* = (L/4\pi R^2 \sigma)^{1/4}$. Thus for the rough fit to the \citet{semenov03a} opacity model given by equation (\ref{eq:opacity}), our estimates of the key dimensionless parameters near the surface become
\begin{eqnarray}
\label{eq:ftcluster}
\tau_* & = & 2.6 f_{g,0.5} f_{*,0.5}^{1/2} \Psi_3^{1/2} \Sigma_1^{3/2} \\
\fE & = & 0.079 f_{*,0.5}^{3/2} \Psi_3^{3/2} \Sigma_1^{1/2}
\label{eq:ftcluster1}
\end{eqnarray}
where $f_{*,0.5}=f_*/0.5$, $f_{g,0.5} = (1-f_*)/0.5$, $\Sigma_1=\Sigma/10$ g cm$^{-2}$, and $\Psi_3 = \Psi/10^3\,\lsun/\msun$; for a zero-age stellar population, $\Psi = 1140 \lsun/\msun$ \citep{fall10a}, so our normalization should be appropriate for a young cluster. In computing $\tau_*$, we have assumed that the gas is arranged in a thin shell (as it must be if it is to be ejected), so $\tau_* = \kappa_* M/4\pi R^2 = \kappa_* \Sigma/4$, with the factor of 4 arising from the difference between the surface density $\Sigma$ as measured by an external observer (which is what appears in the above equations) and the surface density as seen by radiation escaping from the center of the star cluster; if we instead adopt a uniform density sphere geometry, $\tau_*$ will be larger by a factor of 3. Similarly, variations in the dust opacity per unit gas mass could plausibly increase $\tau_*$ and $\fE$ by as much as a factor of a few, and downward by much larger factors in low-metallicity systems. Note that, as pointed out by \citet{fall10a}, $M$ and $R$ enter only through the combination $\Sigma$. The value to which we have normalized $\Sigma$, 10 g cm$^{-2}$, is roughly the maximum observed value for stellar systems anywhere in the Universe \citep{hopkins10a}, and thus the values of $\tau_*$ and $\fE$ above should be regarded as upper limits. Note further that, even though $\fE < 1$ in equation (\ref{eq:ftcluster1}), this does not imply that radiation is unimportant to the dynamics of the dusty gas.  Indeed, radiation pressure should drive strong turbulence as in our simulations presented in Paper I.

We can perform a similar calculation for a galactic disk with total surface density $\Sigma$ and gas mass fraction $f_g$. Such as disk has a surface gravitational force $g=2\pi G \Sigma$. For a stellar population older than $\sim 4$ Myr, the light to star formation rate ratio approaches a roughly constant value $\Phi = 6.1\times 10^{17}\mbox{ erg g}^{-1}\mbox{ s}^{-1} = 1.0\times 10^{10}\,\lsun/(\msun\mbox{ yr}^{-1})$ \citep[calculated using starburst99 --][]{leitherer99a, krumholz07e, krumholz10b}. It is therefore convenient to write the radiative flux as $F = \Phi \dot{\Sigma}_*$, where $\dot{\Sigma}_*$ is the star formation rate per unit area. Plugging this flux into our scalings for the dimensionless parameters gives
\begin{eqnarray}
\label{eq:ftdisk}
\tau_* & = & 0.67 f_{g,0.5} \Phi_{10}^{1/2} \dot{\Sigma}_{*,3}^{1/2} \Sigma_0 \\
\fE & = & 0.43 \Phi_{10}^{3/2} \dot{\Sigma}_{*,3}^{3/2} \Sigma_0^{-1} 
\end{eqnarray}
where $\Sigma_0 = \Sigma/1$ g cm$^{-2}$, $\dot{\Sigma}_{*,3} = \dot{\Sigma}_*/10^3\,\msun$ pc$^{-2}$ Myr$^{-1}$, and $\Phi_{10} = \Phi/10^{10}$ $\lsun/\msun$. The normalizations of $\Sigma$ and $\dot{\Sigma}_*$ here have been chosen to match those of the most vigorously star-forming galaxies observed. Indeed, none of the galaxies in the large sample compiled by \citet{krumholz12a} exceed this star formation rate. Since observations indicate that $\dot{\Sigma}_* \propto \Sigma^p$ with $p\approx 1 - 1.5$, using a normalization for galaxies of lower star formation rates and surface densities would lead to lower values of $\tau_*$ and $\fE$. Thus the values above are, as in the case of single clusters, upper limits for star-forming systems. However, we note that QSO disks on $\sim 1-50$ pc scales can and do exceed these limits \citep{sirko03a, thompson05a}. 

Given these numbers for star clusters and galactic disks, we can draw a few conclusions. The first, already suggested in Paper I, is that, in the absence of additional forces, the dust-reprocessed radiation field cannot launch winds or eject mass from the great majority of star clusters and galaxies. This is because we find that winds are only launched when $\fE > 1$. For star clusters even our upper limit is well below this value, and for galaxies only the most extreme systems approach it, while galactic winds are inferred to be ubiquitous \citep[e.g.][]{veilleux05a}. This is not to say that radiation pressure is not important. As discussed above, if the radiation force is sufficiently strong and $\tau_* < 1$, it may be possible for the \textit{direct} radiation field to eject matter, particularly in systems where gravity is already partially offset by magnetic fields, turbulent motions, or some other force \citep[e.g.][]{murray05a, krumholz09d, murray10a, fall10a, hopkins11a}. Indeed, \citet{krumholz09d} compile a sample of super-star clusters, and show that for some of them the direct radiation force, combined with the momentum of line-driven winds, is likely to be able to eject matter even without significant radiative trapping. Even in somewhat lower luminosity systems where it cannot eject the bulk of the gas, radiation pressure may still be able to drive small amounts of mass to speeds above the escape speed and eject it, as happens for example with massive stars. Nonetheless, our results show that ejecting mass from star-forming systems via radiation pressure is significantly more difficult than many models assume.

A second implication of our work is that, if radiation pressure does launch winds, and if it were the sole driving mechanism, those winds are not likely to carry a momentum flux much larger than a few times $L/c$. Equations (\ref{eq:ftcluster}) and (\ref{eq:ftdisk}) show that $\tau_*$ on galactic scales never much exceeds unity, and that even for the densest clusters it is $<10$; a more typical value for massive clusters would give $\tau_* \sim 1$. Since we find that $\ft \approx 0.5 \tau_*$, this means that we cannot expect winds accelerated primarily by radiation to have $\ft$ larger than $\sim 1$. Thus radiation pressure-driven winds from star-forming systems should not carry a momentum flux that exceeds $L/c$ by more than a few tens of percent. The best fit values of the momentum fluxes of the winds produced by giant star-forming clumps at $z\sim 2$ exceed this limit \citep{genzel11a, newman12b}, which taken at face value would suggest that they cannot be primarily radiation-driven (consistent with \citealt{krumholz10b}). However, we caution that there are very significant uncertainties on these measurements, and for most sources any reasonable estimate of the error bars does not exclude a momentum flux that is close to $L/c$. Moreover, since the winds we observe now were launched some time ago, it is entirely possible that the present-day luminosity we measure for these sources is smaller than it was when the winds were launched. Given the observational uncertainties, we cannot conclude that giant clump winds cannot be radiatively driven, only that, if they are, either their momentum fluxes must have been overestimated or their luminosities at the point of wind launching underestimated.

It is interesting to ask how our conclusions compare with those of prior authors, and why they differ. Most notably, our conclusion that dust-reprocessed radiation is unlikely to be a significant factor in launching winds or disrupting massive clusters is inconsistent with those of a number of authors, including  \citet{murray10a}, \citet{murray11a}, \citet{hopkins11a, hopkins12b}, and \citet{genel12a}. We can understand the difference by examining how the radiation force imparted to the matter is computed in these models versus in our simulations. \citet{murray10a, murray11a} treat the matter as a one-dimensional thin shell, compute the resulting gas temperature and thus the opacity, and compute the radiation force by multiplying the flux by that opacity. Similarly, \citet{hopkins11a, hopkins12b} adopt a uniform, high dust opacity ($\kappa=5$ cm$^2$ g$^{-1}$) and compute the radiation force by multiplying that opacity by the total radiation flux. \citet{genel12a} use a subgrid model that does not explicitly account for radiation forces or any other feedback effect, but manually injects winds at a strength that is tuned to match the results of the \citeauthor{murray10a}~and \citeauthor{hopkins11a}~models. Our numerical results show that these approaches can produce a large overestimate of the radiation force and the trapping factor. \red{In light of this finding,} it seems necessary to recompute the models of \citet{murray10a}, \citet{murray11a}, \citet{hopkins11a, hopkins12b}, and \citet{genel12a} using the approximate fitting formula for $\ft$ that we have derived.

\subsection{\red{Physical Origins of Weak Trapping}}

\begin{figure}
\includegraphics[width=84mm]{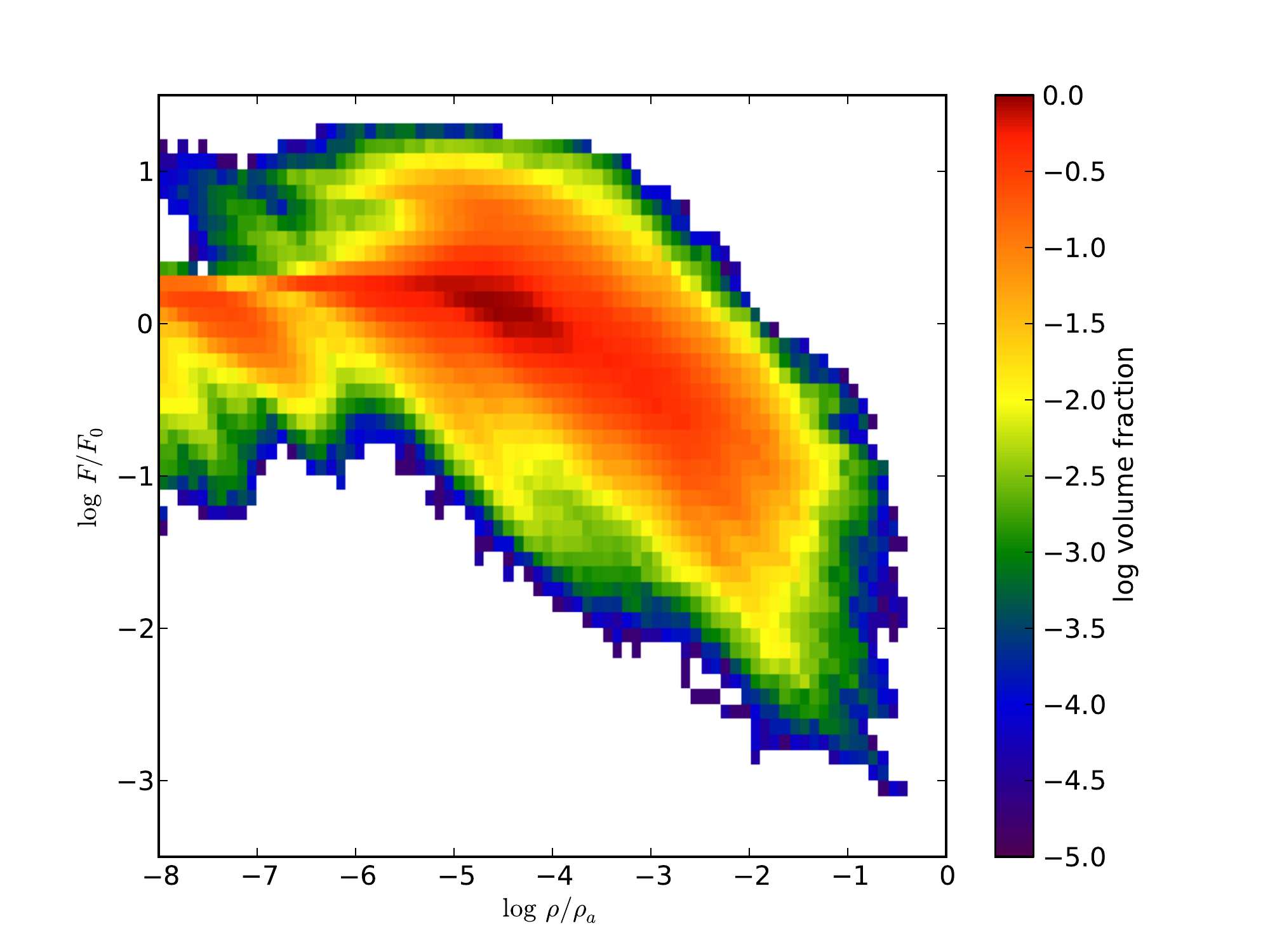}
\caption{
\label{fig:rhof}
\red{
Distribution of gas density and radiation flux throughout the computational volume, for run T10LR at $t=12.4 t_a$, the same run and time shown in Figure \ref{fig:rhotvec}. The color in each pixel of $(\rho/\rho_a, F/F_0)$ indicates the fraction of the simulation volume has the indicated density and total radiation flux. The color scale is normalized so that the value in the most populated bin is unity. White pixels correspond to values of density and flux that are not populated.
}
}
\end{figure}

\red{
Our simulations show that simple estimates of radiative trapping in dusty gas, such as those adopted by many previous authors, tend to produce values of the asymptotic momentum of the dusty gas that are substantially too large. We can identify two significant effects that contribute to this error.
}
First, these models did not take into account the anti-correlation between radiation flux and gas density produced by RRTI, which causes the flux seen by the bulk of the matter to be significantly smaller than the volume-averaged flux. \red{We illustrate this effect in Figure \ref{fig:rhof}, which shows the distribution of density and flux in the simulation volume for a particular time in run T10LR. From the figure, it is clear that dense matter generally has a much lower radiation flux passing through it than more diffuse matter.}
Second, they miss the effect that, because the flux is low within the bulk of the matter, the gas and radiation temperatures are also low. This reduces the opacity within the bulk of the matter, further weakening the matter radiation-coupling.

\red{
It is interesting to investigate further the relative importance of these two effects, particularly because this has implications for how general our results should be. The flux-density anti-correlation would seem likely to be quite independent of the particular opacity law, and thus potentially important even in systems where the opacity has a different functional form (e.g.~electron scattering opacity, which is independent of temperature). The reduction in opacity due to a reduction of the temperature in the bulk of the matter, on the other hand, relies on the particular property that the opacity rises with temperature in dusty gas, and it is unclear how this effect might change for different opacity laws.
}

\red{
To address this question, we compute the quantity
\begin{equation}
\label{eq:taudef}
\langle \tau\rangle \equiv \langle \kappa_R \rho\rangle L_z
\end{equation}
from our simulations. Physically, this is the optical depth computed using the mass-weighted mean opacity. If the flux were uniform (i.e.~$F_z = F_0$ everywhere), it is easy to show that we would have $\ft = \langle\tau\rangle$. Thus the value of $\langle\tau\rangle$ provides some insight into the relative importance of the non-uniformity of the flux. If the flux-density anti-correlation is the dominant effect in reducing $\ft$, since $\langle\tau\rangle$ excludes this effect we should find $\langle\tau\rangle \sim \tau_{\rm IR}$, where $\tau_{\rm IR}$ is the expected trapping factor for a laminar medium as estimated for example from the models of \citet{thompson05a}, \citet{murray10a} and \citet{murray11a}. On the other hand, if $\ft$ is reduced primarily due to the way RRTI reduces the temperature and thus the opacity, then we would expect $\langle\tau\rangle\sim \ft$ instead.
}

\begin{figure}
\includegraphics[width=84mm]{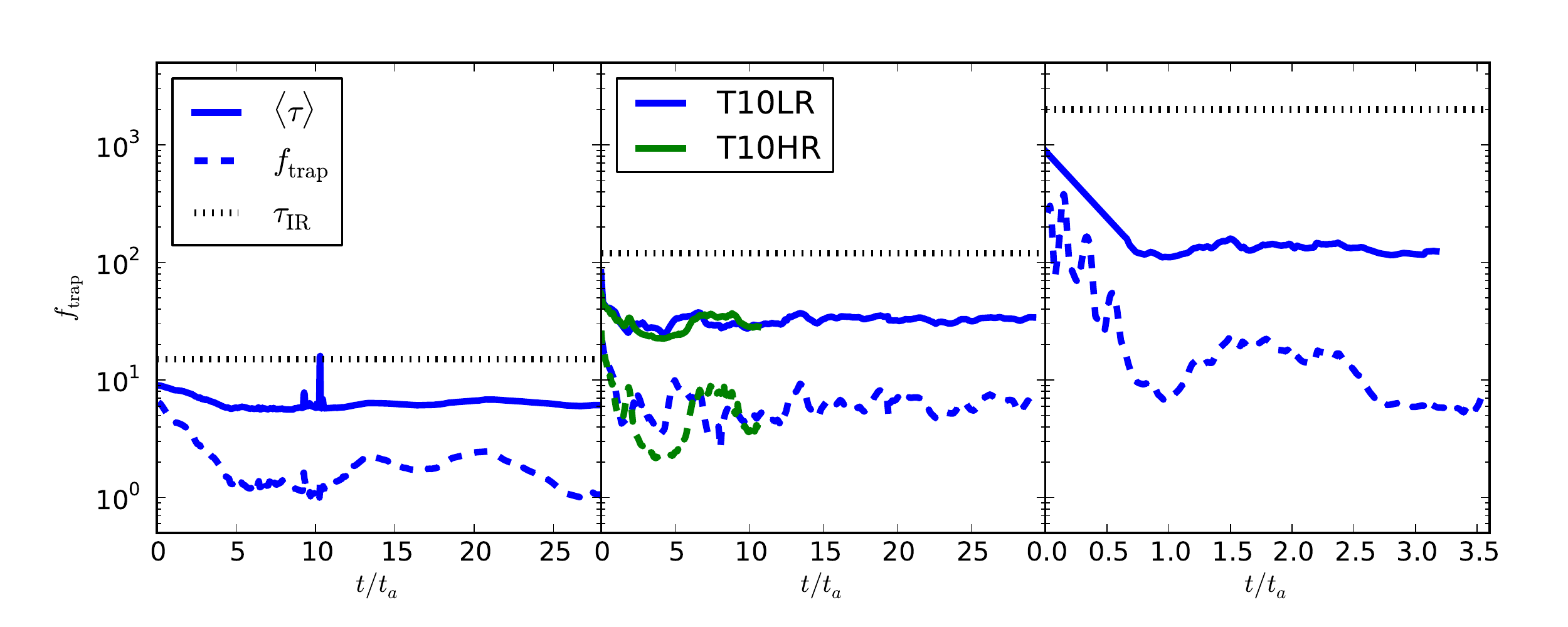}
\caption{
\label{fig:odepth}
\red{
Values of $\langle\tau\rangle$ as a function of $t$ for all simulations (solid lines). For comparison, we also show $\tau_{\rm IR}$ (dotted lines) and $\ft$ (dashed lines; identical to the values shown in the lower panels of Figure \ref{fig:ftrap}). 
}
}
\end{figure}

\red{
We show $\langle\tau\rangle$ as a function of time in Figure \ref{fig:odepth}, where for comparison we also plot $\ft$ and $\tau_{\rm IR}$. We also report the time-averaged value of $\langle\tau\rangle$, computed in the same manner as for $\ft$, in Table \ref{tab:simresult}.\footnote{\red{Careful readers may notice that, for run T30, the plot of $\langle\tau\rangle$ does not extend quite as far in time as the plot of $\ft$, and in some places is more smoothly interpolated in time; this is due to a coding error that caused some portion of the output from this run to be lost, preventing us from performing the post processing required to calculate $\langle\tau\rangle$ at the corresponding times. This issue did not affect $\ft$ because the values required for it were calculated as the simulation ran.}} We find that $\langle\tau\rangle$ is, in a logarithmic sense, roughly half-way between $\tau_{\rm IR}$ and the actual value of $\ft$ we measure in our simulations. This indicates that the flux-density anti-correlation and the reduction in opacity due to non-uniform temperatures are about equally important in reducing the rate of momentum transfer from radiation to gas in our dusty flows. That there is at least some reduction in the momentum transfer to the gas even due to flux-density anti-correlation even when the opacity is gray is qualitatively consistent with the findings of \citet{jiang13a}, who find RRTI operating in a medium with a gray opacity.
}

\red{
Based on this result, we tentatively predict that even for an opacity that depends on a different positive power of the temperature, the flux-density anti-correlation induced by RRTI should produce some reduction in $\ft$ compared to what one would estimate using a model in which the density and radiation flux are taken to be uniform. However, if the opacity is gray this reduction should be smaller than what we find here by roughly a factor of 2 in a logarithmic sense. These issues are particularly important in the case of dust opacities since it is only for $T < 150-200$ K that $\kappa_R\propto T^2$.  At higher temperatures the dust opacity is approximately constant with temperature \citep{semenov03a}. Thus, in some regimes we would expect the flux-density anti-correlation to dominate. Compared to our $\kappa_R\propto T^2$ model, a dust opacity that flattened at high temperatures would probably produce slightly lower $f_{\rm trap}$ (since the decrease in opacity should at least marginally reduce the net radiation force) but much lower $\tau_{\rm IR}$ (since the opacity at the midplane would be greatly reduced), and thus less of a difference between $\tau_{\rm IR}$ and $f_{\rm trap}$.
}

\red{
It is unclear what would happen in the case of an opacity that falls with temperature, as is the case for a Kramers opacity law, but it is conceivable that in this case $\ft$ might actually be closer to the naive laminar estimate, since for such an opacity the temperature-dependence induced by the non-uniform radiation field would tend to increase rather than decrease the strength of matter-radiation coupling. However, we emphasize that all of these conclusions are tentative, since it is entirely possible that the non-linear development of the RRTI, and thus the amount of flux-density anti-correlation it produces, itself depends on the opacity law.
}

\subsection{Relation to Dusty Star Winds}

Although our work is focused on the problem of star cluster and galactic winds, the general problem of radiative driving of dusty gas also arises in the context of winds from dusty late-type stars \citep[e.g.][]{goldreich76a, habing96a}. Before proceeding, it is important to point out a significant way in which this problem differs from our work here and in Paper I. At the low temperatures typical of interstellar gas even in intensely star-forming galaxies, the opacity roughly $\kappa\propto T^2$ \citep[e.g.][]{semenov03a}, as we use in our models. This means that the opacity general drops monotonically with height in an atmosphere. In the case of dusty stellar winds, on the other hand, where temperatures are near the grain sublimation temperature, the opacity is much more complex and non-monotonic, both due to grain formation, destruction, and drift relative to the gas, and because even for a constant grain population the opacity varies are roughly $T^0$ (i.e.~constant) rather than $T^2$ at temperatures close to the grain sublimation temperature. \red{As discussed in the previous section, for such an opacity law we tentatively expect that there will be some reduction in $\ft$ compared to the expected values for a uniform medium, but that it will be weaker than what we find here.}

With this caveat aside, we note that one-dimensional models have been reasonably successful at reproducing many observations of massive star winds \citep[e.g.][]{ivezic95a, ivezic10a, elitzur01a}, implying that RRT instability may not be critical for these stars. However, the observationally-inferred momenta of dusty star winds are usually below $L/c$ \citep[e.g.][]{groenewegen09a}, and that even the highest inferred momenta are no more than $\sim 10 L/c$ \citep[e.g.][]{elitzur01a}, which in turn suggests that these stars are not in the regime where we require strong amplification of the force by radiative trapping that might be inhibited by RRTI. \red{On the other hand, one might legitimately worry that even the more modest reduction in optical depth we expect for a gray opacity might make it difficult to accelerate winds to $10 L/c$. We tentatively conclude that there might be some tension between our numerical results and the current generation of AGB wind models, and potentially even some tension with the observations. However, the problem clearly needs further investigation, since our estimate of the factor by which RRTI should reduce $\ft$ for complex opacity laws is highly uncertain.}

\subsection{Limitations Due to Geometric Simplifications of the Simulations}

Our simulations represent an idealized numerical experiment with a simple geometry. It is therefore interesting to ask how a more realistic setup would likely affect our results. One obvious simplification in our simulations is that they are two- rather than three-dimensional. The implications of this are discussed extensively in Paper I, and we refer readers to the discussion there.

A second simplification is that we have assumed a planar geometry, whereas a real wind will generally approach a spherical geometry, at least once it is far from its launch point. A small section of a spherical shell of wind material may be treated as planar, and so our planar results should continue apply locally. The main difference between planar and spherical geometries, therefore, is that in planar geometry $\tau_*$ and $\fE$ are fixed, whereas for a spherical wind they will vary as the wind expands. This variation is caused by two effects. First, as a spherical shell of constant mass expands in radius $R$, its surface density drops as $R^{-2}$, which reduces $\tau_*$. Second, the gravitational force $g$ and the flux $F_0$ encountering a spherical shell also both drop as $R^{-2}$. This means that the ratio $F_0/g$ remains constant; however, the drop in $F_0$ reduces $T_*$ as $R^{-1/2}$ and thus $\kappa_{R,*}$ as $R^{-1}$, which affects both $\fE$ and $\tau_*$. The net effect is that, for a spherical shell of fixed mass and radius $R$, $\fE\propto R^{-1}$ and $\tau_* \propto R^{-3}$, and thus an expanding spherical shell traces a line of slope $3$ in the $(\log\fE, \log\tau_*)$ plane depicted in Figure \ref{fig:plotfit}. Systems start at the upper right of the plane, then move down and to the left as they expand. Examining the Figure, we see that such a trajectory will result in a value of $\ft$ that decreases with time, and that it approaches an asymptotically constant value of $\fEavg$. This is not surprising; it is simply a statement that, as a shell expands, its optical depth drops and thus the dust-reprocessed radiation field becomes less and less important compared to the direct one, which gives constant $\fEavg$.

A third simplification of our simulations is that we have assumed a constant flux as would would be produced in a galaxy with all the stars at the midplane, or by a star cluster with all the stars concentrated at the center and the mass at fixed radius. In reality, the sources of radiation are intermixed with the gas being launched in a wind. This may well result in a significant reduction of the direct radiation force due to geometric cancellations. However, it should not substantially affect our results for $\ft$, simply because an appreciable value of $\ft$ requires that the radiation field be trapped and therefore isotropized, forgetting its original direction. Thus our results for $\ft$ should be robust against a change from planar or point-like sources to distributed sources.

\subsection{\red{Limitations of the Flux-Limited Diffusion Approximation}}
\label{sec:fld}

\red{
Our simulations make use of the 2TFLD approximation, which is an approximate treatment of radiative transfer. The approximation is highly accurate when the optical depth is very high, but at low optical depth is loses information about the directionality of the radiation field, leading to an inaccurate treatment of shadowing and similar effects. It is therefore important to ask to what extent our results might be affected by the limitations of our numerical method. While a full answer to this question can only come from repetition of the simulation with a more accurate radiation transport method, we can make a few general observations here.
}

\begin{figure}
\includegraphics[width=84mm]{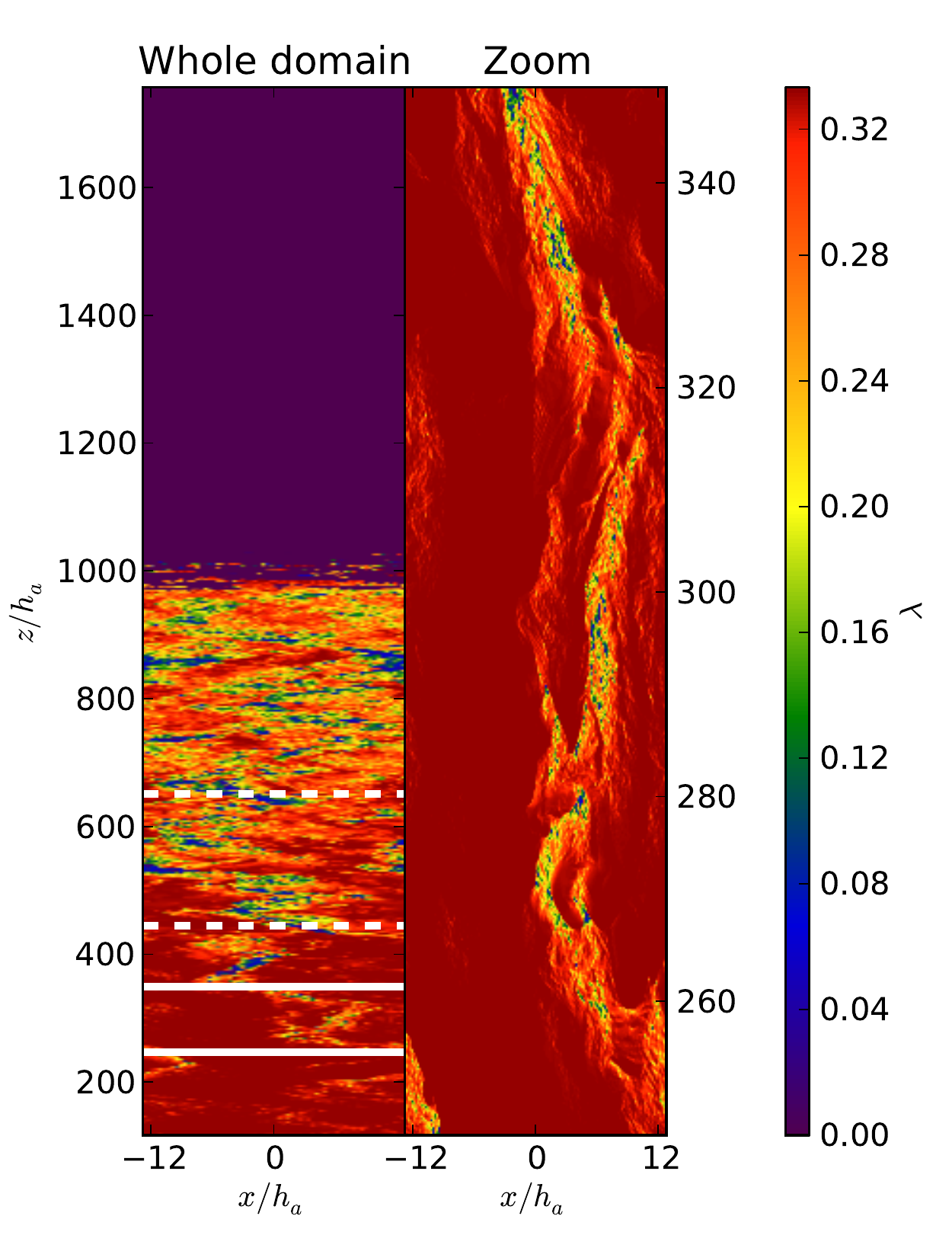}
\caption{
\label{fig:limiter}
\red{
Values of the flux limiter $\lambda$ at the same time and for the same run as shown as in Figure \ref{fig:rhotvec}: $t=12.4t_a$ in run T10LR. Values of $\lambda=1/3$ indicate optically thick regions, while values near 0 indicate optically thin ones. The left panel shows the entire computational domain, while the right panel shows the same zoomed-in region as in Figure \ref{fig:rhotvec}, which is centered on the vertical center of mass. Note that, in the left panel, the $x$ direction has been stretched by a factor of 16 relative to the $z$ direction in order to render the image readable. In the left panel, the white horizontal solid lines indicate the upper and lower boundaries of the zoomed-in region on the right. The two white horizontal dashed line indicates the values of $z$ below which 90\% and 99\% of the mass lie.
}
}
\end{figure}

\red{
Figure \ref{fig:limiter} shows the flux limiter $\lambda$ in one of our calculations at the same time as shown in Figure \ref{fig:rhotvec}. Values of $\lambda$ close to $1/3$ correspond to locations where the optical depth is high, and the 2TFLD approximation should be very accurate, while values of $\lambda$ close to $0$ correspond to optically thin regions. The white horizontal dashed lines indicate the values of $z$ below which 90\% and 99\% of the mass in the computational domain lie. From the figure, it is clear that the great majority of the mass lies in a region where $\lambda$ is very close to $1/3$. There are smaller values of $\lambda$ within the channels through which radiation is escaping, but even there $\lambda$ only falls to $\sim 0.25-0.3$ except in tiny volumes. It seems that the channels that carry most of the flux are optically thin only in a relative sense, i.e.~they have much lower optical depth than most of the higher-density gas around them, but they still have optical depths $\sim 1$, not $\ll 1$. In contrast, $\lambda$ does approach 0, as expected, high above where most of the matter is located.
}

\red{
Based on this figure, it seems likely that our treatment of radiation transport is reasonably accurate for the great majority of the mass. Where one might legitimately be concerned with the accuracy of the 2TFLD approximation is in the intermediate optical depth regime that lies, roughly, in the range from $z/h_a \approx 500-1000$ in the plot. However, as the dashed horizontal lines indicate, this regions contains no more than a few percent of the mass in the simulation domain. In essence, we find that almost all the gas is in an optically thick shell where $\lambda\approx 1/3$ and the 2TFLD approximation should be accurate, while a very small fraction is in a very extended atmosphere where its accuracy might be poor. However, since we are primarily concerned with the momentum imparted to the bulk of the mass, this extended atmosphere should have little effect on our main result, which is that this momentum is quite limited. Nonetheless, we caution that the precise details of the structure of the atmosphere, or the interiors of the low optical depth channels further within the shell, might be affected by the limitations of the 2TFLD method. These should therefore be treated with some caution.
}

\red{
Finally, we note that \citet{jiang13a} have investigated the problem of RRTI in an optically thick medium for the case where the opacity is dominated by electron scattering rather than dust absorption, and have done so using a variable Eddington tensor (VET) method that should be more accurate than 2TFLD. They compare their VET results to those obtained using Eddington approximation, which is closely related to the diffusion approximation we use. The results of their comparison are consistent with our conclusions in this section. They find that the primary effect of going from the Eddington approximation to VET is to suppress some of the small-scale structure induced by RRTI, leading to larger-scale filaments and channels (their Figure 9). This also has the effect of delaying the initial onset of instability some, because in the linear regime small modes grow fastest, and suppression of such modes leads to somewhat slower growth at first. However, once larger wavelength modes go non-linear, the difference in growth rates is small (their Figure 8). Since the time spent during the linear phase of instability growth for small modes is astrophysically negligible for the parameters describing real ULIRGs and massive star clusters (see the discussion in Paper I), it seems unlikely that a more accurate radiation transport method would change our results qualitatively.
}

\section{Summary}
\label{sec:summary}

In this paper we analyze the properties of optically thick radiation pressure-driven dusty winds. We consider the idealized problem of a column of material through which a specified radiation flux is passed. We first show that such a system is characterized by a single dimensionless number, $\tau_*$, the optical depth of the matter computed using the opacity at the dust photosphere, and that this parameter will determine the rate at which the matter column absorbs momentum from the radiation field. We then use radiation-hydrodynamic simulations to measure this momentum transfer rate. We find that, after one to a few dynamical times, radiation Rayleigh-Taylor instability (RRTI) drives the gas into a configuration where most of the matter is in dense filaments aligned along the direction of the radiation flux, while most of the radiation flux passes through channels of reduced optical depth between the filaments. This configuration minimizes matter-radiation interaction, and thus limits the rate at which matter can take up momentum from the radiation field.

We combine this result with the result from Paper I, where we considered irradiated columns of matter confined by gravity, to produce a fitting formula for the behavior of irradiated, gravity-confined dusty gas layers. The behavior of these structures is determined by $\tau_*$ and by $\fE$, the ratio of radiative and gravitational forces at the dust photosphere. We identify three regimes of behavior depending on the values of these parameters. At a given $\tau_*$, there is a critical value of $\fE$ below which radiation is dynamically unimportant. For values of $\fE$ above the critical value but below unity, RRTI sets in and makes the gas turbulent, but does not produce a wind. Only for $\fE \ga 1$  is there a wind, and even in the limit where gravity provides negligible confinement of that wind, the wind momentum flux is roughly $1 + 0.5 \tau_*$ times the radiation momentum flux.

We then consider the implications of these results for star-forming clusters and galaxies. For observed clusters and galaxies, our results suggest that dust-reprocessed radiation is unlikely to be able to drive winds and eject matter. The direct radiation field may still be able to launch winds, but only in systems where its momentum alone is sufficient to overcome gravity, without significant amplification by radiative trapping.

\section*{Acknowledgements}

\red{We thank the anonymous referee for a helpful report.} MRK acknowledges support from the Alfred P.~Sloan Foundation,  the NSF through grant CAREER-0955300, and NASA through Astrophysics Theory and Fundamental Physics Grant NNX09AK31G, and a Chandra Space Telescope Grant.
TAT acknowledges support from the Alfred P.~Sloan Foundation and NASA grant NNX10AD01G.
This research was supported by allocations of advanced computing provided by the National Science Foundation and by the NASA High-End Computing Program. Simulations were performed on Kraken at the National Institute for Computational Sciences (http://www.nics.tennessee.edu/), Ranger at the Texas Advanced Computing Center at The University of Texas at Austin (http://www.tacc.utexas.edu), and Pleiades at the NASA Advanced Supercomputing (NAS) Division at Ames Research Center.

\begin{appendix}

\section{\red{Temperature and Dynamical Approximations}}
\label{sec:assumptions}

\red{Here we justify three physical approximations we have made in our formulation of the governing equations: first, that the dust and gas temperatures are tightly coupled; second, that, at least in regions of high opacity, the dust and radiation temperatures are as well; third, that the dust and gas are dynamically coupled and can be treated as a simple fluid.
}

\subsection{\red{Thermal Behavior}}

\red{
To check the first two of these assumptions, we must compare the rate of dust-gas energy exchange via collisions with other rates of heating and cooling in the problem. Specifically, we must check that (1) the rate of energy exchange between the dust and the radiation field greatly exceeds the rate of collisional energy exchange between the dust and gas, ensuring that the dust is always well-coupled the radiation field, and (2) the rate of collisional energy exchange between the dust and gas is much larger than any other rate of heating or cooling for the gas alone, so that the gas is forced to the dust temperature. In what follows, unless stated otherwise, the rates of all heating and cooling processes follow the approximations described in \citet{krumholz13b}.}

\red{The rate of dust-gas energy exchange per unit volume is
\begin{eqnarray}
\Psi_{\rm gd} & = & \alpha_{\rm gd} (\rho/\mu_{\rm H})^2 T_g^{1/2} (T_d - T_g) \\
& = & 5.8\times 10^{-16} \rho_{-16}^2 T_{g,2}^{1/2} (T_d-T_g)_2 \mbox{ erg cm}^{-3}\mbox{ s}^{-1}
\end{eqnarray}
where $\alpha_{\rm gd} \approx 3.2\times 10^{-34}$ erg cm$^3$ K$^{-3/2}$ is the dust-gas collisional coupling coefficient, $\mu_{\rm H}\approx 2.3\times 10^{-24}$ g is the gas mass per H nucleus, $T_g$ is the gas temperature, $T_d$ is the dust temperature, $\rho_{-16} = \rho/10^{-16}$ g cm$^{-3}$, $T_{g,2} = T_g/100$ K, and $(T_d-T_g)_2=(T_d-T_g)/100$ K. The scalings we have chosen here for $\rho$ and $T$ are intentionally close to $\rho_a$ and $T_*$, so the numerical coefficients should be roughly representative of the actual values in the simulation, at least during the initial phases. The value to which we scale $T_d-T_g$ is intended to be representative of what would be required to generate a relative dust-gas temperature difference of order unity. The rate of dust energy exchange with the radiation field per unit volume is
\begin{eqnarray}
\Psi_{\rm dr} & = & \kappa_P \rho (4\pi B - c E) = \kappa_P \rho c a (T_d^4 - T_r^4) \\
& = & 2.3\times 10^{-13} \rho_{-16} (T_d-T_r)_2^4  \mbox{ erg cm}^{-3}\mbox{ s}^{-1}
\end{eqnarray}
where we have written $E = a T_r^4$ for convenience.
}

\red{
For the gas, the competing heating processes are cosmic ray heating and shock heating, while the competing cooling process is molecular line emission. The cosmic ray heating rate is
\begin{eqnarray}
\Gamma_{\rm CR} & = & \zeta q_{\rm ion} \rho/\mu_{\rm H} \\
& = & 1.0\times 10^{-17} \zeta_{-14} \rho_{-16} \mbox{ erg cm}^{-3}\mbox{ s}^{-1}
\end{eqnarray}
where $\zeta$ is the primary ionization rate per H nucleus from cosmic rays, $\zeta_{-14} = \zeta/10^{-14}$ s$^{-1}$, and $q_{\rm ion} \approx 15$ eV is the thermal energy added per primary ionization. The cosmic ray ionization rate to which we have scaled here is roughly 100 times a typical Milky Way value \citep[e.g.][]{indriolo12a}, consistent with the values predicted by \citet{lacki10a} for starburst galaxies. There is also direct evidence for enhanced cosmic ray fluxes in starburst galaxies from $\gamma$-ray emission \citep{abdo10a}. It is conceivable that the cosmic ray flux could be even higher than this, which would have important implications of this for the thermodynamics of the gas in ULIRGs, see \citet{narayanan11a, narayanan12a}. The shock heating rate of course varies with position and time, but we can obtain a crude order of magnitude estimate by positing that the full kinetic energy per unit volume of the gas is dissipated into shocks each crossing time. This implies a heating rate
\begin{eqnarray}
\Gamma_{\rm shock} & \approx & \rho \sigma^3 / h_g \\
& = & 3.2\times 10^{-17} \rho_{-16} \sigma_6^3 h_{g,0}^{-1} \mbox{ erg cm}^{-3}\mbox{ s}^{-1}
\end{eqnarray}
where $\sigma$ is the gas velocity dispersion and $h_g$ is the characteristic size of the region occupied by the gas and $h_{g,0} = h_g/1$ pc. The scaling here is again chosen to be close the the characteristic size of the structures seen in our simulations, which is $\sim 10-100h_a$. Finally, the rate of molecular line cooling is the most difficult to estimate, since it depends on the complex physics of optically thick line emission. We use the \textsc{despotic} package \citep{krumholz13b} to compute the cooling rate via CO emission for a cloud of volume density $10^{-16}$ g cm$^{-3}$, column density $\Sigma = 5$ g cm$^{-2}$, and gas temperature $100$ K, with a CO abundance of $1\times 10^{-4}$ per H nucleus, and obtain
\begin{equation}
\Lambda_{\rm CO} = 3.5\times 10^{-17}  \mbox{ erg cm}^{-3}\mbox{ s}^{-1}.
\end{equation}
}

\red{
Based on these calculations, we see that, for material with density $\sim \rho_a$, temperature $\sim T_*$, and velocity dispersion $\sim 10$ km s$^{-1}$ (comparable to what we find in our simulations), we expect $|\Psi_{\rm dr}| \gg |\Psi_{\rm g}| \gg |\Gamma_{\rm CR}| + |\Gamma_{\rm shock}| + |\Lambda_{\rm CO}|$. The former inequality ensures that the dust and radiation field will be forced to the same temperature, while the latter ensures that the dust and gas will be forced to the same temperature.   Our result is not surprising in light of earlier work. \citet{goldsmith01a} shows that dust and gas as well-coupled once the density exceeds $\sim 10^4 - 10^5$ cm$^{-3}$, and \citet{narayanan11a, narayanan12a} conclude that strong dust-gas coupling dominates the thermodynamics of gas in ULIRGs.
}

\red{
However, we do caution that the gap between the dust-gas energy exchange rate and the dominant gas heating and cooling terms is only a factor of $\sim 10$. At late times in the simulations, the density drops significantly as material spreads out, and once this happens the gas and dust should thermally decouple. Since the gap between the gas-dust energy exchange rate and the dust-radiation energy exchange rate is much larger, the dust would remain thermally coupled to the radiation field, but the gas would not. In practice this means that the error in our assumption that the gas is at the same temperature as the dust manifests not as an error in the opacity (which would still be determined by the coupled dust and radiation temperatures, consistent with our assumptions), but instead as an error in the gas thermal pressure. This is unlikely to affect the dynamics significantly, because the gas velocities are generally highly supersonic, and thus thermal pressure is weak compared to ram pressure.
}

\subsection{\red{Dynamical Behavior}}

\red{
We now investigate the final assumption, which is that dust and gas are dynamically-well coupled, and do not drift significantly relative to one another. If the only mechanism capable of exchanging momentum between gas and grains is collisions, then the characteristic free-streaming length of a dust grain of mass $m_D$ with respect to the gas is
\begin{equation}
\lambda_{\rm dg} \approx \frac{m_D}{\mu} \frac{1}{n_g \sigma},
\end{equation}
where $\mu$ is the mean mass per gas particle, $n_g$ is the number density of gas particles, and $\sigma$ is the cross-section for grain-gas collisions. This equation holds as long as the grain is small compared to the particle mean-free path, which will be the case for typical interstellar grains. If we take the grain to be a sphere of radius $a$ and density $\rho_D$, and the cross section $\sigma = \pi a^2$, then we can rewrite this equation as
\begin{eqnarray}
\label{eq:lambdadg}
\lambda_{\rm dg} & \approx & \frac{4}{3} a \frac{\rho_D}{\rho} \\
& = & 1.3\times 10^{-7} a_{-1} \rho_{D,3} \rho_{-16}\mbox{ pc},
\end{eqnarray}
where $a_{-1} = a/0.1$ $\mu$m, $\rho_{D,3} = \rho_D/3$ g cm$^{-3}$, and the values to which we have scaled in the numerical evaluation are typical for interstellar silicate grains.
}

\red{
In comparison, the characteristic length scale for our problem is $h_a \sim 10^{-2}$ pc, and thus the characteristic grain streaming length is a factor of $\sim 10^5$ smaller. We can therefore safely ignore grain streaming at the stages of our simulation where the structure is developing. As with thermal coupling, at very late times once the density drops by $\sim 5$ orders of magnitude, grain streaming with respect to the matter can become significant, and the approximations we use may begin to break down. However, we note that equation (\ref{eq:lambdadg}) represents a maximum possible grain streaming length, because it ignores magnetic coupling between grains and gas. If the grains are charged, then magnetic forces will tie them to the gas more strongly than hydrodynamic collisions, and make free-streaming less important.
}

\section{Initial Conditions for Run T30}

For run T30, we do not have initial conditions from Paper I because we did not perform any runs with $\tau_* = 30$. To generate such conditions, we run a simulation with $\tau_* = 30$ and $\fE = 0.05$ (i.e.\ with gravity turned on) following the same procedure as for all other runs described in Paper I. We refer to this run as T30F0.05. As in the other runs from Paper I, we perform the simulation at a resolution $\Delta x = 0.5 h_*$, in a computational domain of $1024\times 16384$ cells, corresponding to a size of $512h_* \times 8192h_*$. We run the simulation for a time $t = 75t_*$, by which point a turbulent flow is fully developed. To produce initial conditions for run T30 in this paper, we must rescale the results of run T30F0.05, because $h_a/h_* = 600$, so that the resolution of run T30F0.05 is $\Delta x = 8.3\times 10^{-4} h_a$. This is so high that it would be impossible to advance the run for a time comparable to $t_a$. We therefore downsample the output at the final time in run  T30F0.05 by a factor of 32, producing a resolution of $\Delta x = 0.027 h_a$. We also replicate the density, velocity, gas temperature, and radiation energy density fields 32 times in the horizontal direction; since run T30F0.05 has periodic boundary conditions, this is fully self-consistent. The result is a cube of initial conditions that is $1024\times 512$ cells in size, at a resolution $\Delta x = 0.027 h_a$, corresponding to a physical size $27.3 h_a \times 13.6 h_a$. We use this state as the initial condition for run T30, extending the computational domain in the vertical direction exactly as for the other runs described in Section \ref{sec:sim}.

\end{appendix}

\bibliographystyle{mn2e}
\bibliography{refs}

\end{document}